\documentclass[11pt]{article}
\pdfoutput=1
\usepackage{bbold}
\usepackage{amsmath}
\usepackage{url}
\usepackage{hyperref}
\usepackage{slashed}
\usepackage{tikz}
\usetikzlibrary{decorations.pathmorphing}
\usetikzlibrary{matrix}
\usepackage{graphicx}
\usepackage{epstopdf}
\usepackage{subfigure}
\usepackage{pgfplots}
\usepackage{amssymb}
\usepackage{color}
\usepackage{mathrsfs}
\usepackage{fancybox}
\usepackage{lipsum}
\usepackage{eurosym}
\usepackage{tcolorbox}
\usepackage{anyfontsize}
\usepackage{enumitem,kantlipsum}

\usetikzlibrary{decorations.pathmorphing}
\tikzset{snake it/.style={decorate, decoration=snake}}
\let\originallesssim\lesssim
\let\originalgtrsim\gtrsim
\DeclareRobustCommand{\lesssim}{%
  \mathrel{\mathpalette\lowersim\originallesssim}%
}
\DeclareRobustCommand{\gtrsim}{%
  \mathrel{\mathpalette\lowersim\originalgtrsim}%
}

\makeatletter
\newcommand{\lowersim}[2]{%
  \sbox\z@{$#1<$}%
  \raisebox{-1pt}{\small $\m@th#1#2$}%
}
\def\({\left (}
\def\){\right )}
\def\[{\left [}
\def\]{\right ]}

\numberwithin{equation}{section}

\usepackage{environ}
\NewEnviron{myequation}{%
\begin{eqnarray}
\scalebox{1.05}{$\BODY$}
\end{eqnarray}
}

\newcommand{\beq}{\begin{equation}}
\newcommand{\eeq}{\end{equation}}
 
\newcommand{\bea}{\begin{eqnarray}}
\newcommand{\ea}{\end{eqnarray}}
\newcommand{\barr}{\!\begin{array}}
\newcommand{\earr}{\end{array}\!}

\usepackage{mciteplus}

\def\spc{\hspace{1pt}}

\counterwithout{equation}{section}

\def\smpc{{\hspace{.5pt}}}

\usetikzlibrary{decorations.pathreplacing,decorations.markings,snakes}

\def\be{\begin{equation}}
\def\ee{\end{equation}}

\def\la{\langle}
\def\bea{\begin{eqnarray}}
\def\eea{\end{eqnarray}}
\def\is{\!  & \!  = \!  &  \!}
\def\ra{\rangle}

\def\ea{\eea}

\def\qbigrr{\smpc,q\bigr)}

\def\be{\bea}
\def\ee{\eea}

\def\tr{{\rm tr}}

\def\spc{\hspace{1pt}}
\def\is{\! &  \! = \! & \!}

\def\lL{{{\mbox{\fontsize{7pt}{7.5pt}${L}$}}}}
\def\rR{{{\mbox{\fontsize{7pt}{7.5pt}${R}$}}}}
\usepackage{titling} 
\usepackage[affil-it]{authblk} 
\renewcommand{\footnotesize}{\small}
\begin{document}

\addtolength{\topmargin}{3cm}

\title{\bf Double-scaled SYK and de Sitter Holography}

\author[1]{Vladimir Narovlansky,}
\author[1,2]{Herman Verlinde}

\affil[1]{ Department of Physics, Princeton University, Princeton, NJ 08544, USA}

\medskip

\affil[2]{School of Natural Sciences, Institute for Advanced Study, Princeton, NJ 08540 USA}
\date{\small \texttt{narovlansky@princeton.edu, verlinde@princeton.edu}}

    \maketitle
\bigskip

\bigskip

    \begin{abstract}
   {We propose a new model of low dimensional de Sitter holography in the form of a pair of double-scaled SYK models at infinite temperature coupled via an equal energy constraint $H_L=H_R$.  As a test of the duality, we compute the two-point function between two dressed SYK operators ${\cal O}_\Delta$ that preserve the constraint. We find that in the large $N$ limit, the two-point function precisely matches with the Green's function of a massive scalar field of mass squared $m^2 = 4\Delta(1-\Delta)$ in a 3D de Sitter space-time with radius $R_{\smpc\rm dS}/G_N = 4\pi N/p^2$. In this correspondence, the SYK time is identified with the proper time difference between the two operators. We introduce a candidate gravity dual of the doubled SYK model given by a JT/de Sitter gravity model obtained via a circle reduction from 3D Einstein-de Sitter gravity. We comment on the physical meaning of the finite de Sitter temperature and entropy.}
    \end{abstract}

\addtolength{\topmargin}{-3cm}
\pagebreak

\addtolength{\baselineskip}{-0.1mm}
\addtolength{\parskip}{-.1mm}
\tableofcontents
\addtolength{\baselineskip}{0.2mm}
\addtolength{\parskip}{.7mm}
\addtolength{\abovedisplayskip}{.65mm}
\addtolength{\belowdisplayskip}{.65mm}
\pagebreak

\section{Introduction}

\vspace{-1.5mm} 

Holography aims to represent gravitational physics inside a space-time region in terms of a dual quantum system. This framework has been successfully applied to anti-de Sitter space and the near horizon region of black holes, where solvable theories such as ${\cal N}=4$ SYM and more recently the SYK model  \cite{kitaevTalks,Maldacena:2016hyu,Polchinski:2016xgd,Cotler:2016fpe,Berkooz:2018jqr,Berkooz:2018qkz, Lin:2022rbf,  Lin:2023trc} have produced fruitful insights. While it is generally believed that cosmological space-times such as de Sitter space should admit a similar holographic description, progress on this front has been more limited. Here we report on some new results that look like a step forward in this direction.

There are several hints that the high temperature limit of the double scaled SYK (DSSYK) model can provide a quantum description of low-dimensional de Sitter space
\cite{HVtalks}\cite{Susskind:2021esx,Susskind:2022bia, Susskind:2022dfz, Lin:2022nss}\cite{Rahman:2022jsf}.
The first two pieces of evidence are that DSSYK has a natural maximal entropy state and a bounded energy spectrum labeled by an angle $\theta$ \cite{Berkooz:2018qkz}. The same is true on the gravity side: pure de Sitter space is believed to represent a maximal entropy state  and placing a localized mass $M$ in 3D de Sitter space produces a conical singularity with an angle deficit $\theta$ proportional to the mass $M$ measured in Planck units. This relation is further substantiated by the fact that SYK correlation functions exhibit the same quantum group symmetry \cite{Berkooz:2018jqr} that governs the braiding properties of point masses in 3D de Sitter gravity and by the apparent match between the edge theory of 3D de Sitter gravity \cite{Klemm:2002ir} and the bi-local collective field theory of the SYK model.

The above formal correspondences between double scaled SYK and de Sitter gravity were noted some time ago  \cite{HVtalks}. Expanding them into a concrete holographic dictionary, however, remained a non-trivial task for several reasons. Unlike AdS, de Sitter space doesn't have an obvious preferred location for where to place a holographic screen. Proposals include \cite{Strominger:2001pn,  Alishahiha:2004md, Banks:2006rx, Anninos:2011ui, Anninos:2011af, Anninos:2017hhn, Anninos:2017eib, Dong:2018cuv, Coleman:2021nor, Chandrasekaran:2022cip, Witten:2023xze}. The dS/CFT prescription \cite{Strominger:2001pn}, which places the dual CFT on past or future infinity, does not provide a true identification between unitary quantum systems and leaves the interpretation of the de Sitter temperature, entropy, and time evolution rather obscure. Indeed, one of the most compelling arguments and guiding principles for de Sitter holography is the Gibbons and Hawking interpretation of the area of the cosmological horizon bounding the static patch as an entropy \cite{Gibbons:1976ue}. A more promising approach, therefore, as advocated most recently in \cite{Chandrasekaran:2022cip}\cite{Witten:2023xze}, is to try to construct the dual quantum system to de Sitter space via the requirement that it encompasses the operator algebra on the time-like trajectory of a localized observer in de Sitter space. A downside of this perspective is that gravity remains dynamical on the observer world-line, which makes it less evident how to define an algebra of observables in such a fluctuating space-time environment. In dimensions D $\leq$ 3, however, this problem is ameliorated by the fact that there are no local metric degrees of freedom and that pure de Sitter quantum gravity in $D\leq 3$ is soluble \cite{Jackiw:1984je, Witten:1988hc, Witten:1989ip}. 

What should the operator algebra and Hilbert space of a local de Sitter observer look like? 
Semi-classically, and in the absence of external forces, the observer follows a geodesic. It is natural to identify the causal wedge associated with this geodesic with the static patch of de Sitter space. The observer can probe the static patch by emitting and detecting particles by means of local quantum fields. 
In the non-interacting semi-classical limit, the two-point function  
\bea
G_\Delta(\tau_2,\tau_1) = \la \Psi_{\rm dS} | {\cal O}_{\!\Delta} (\tau_2) {\cal O}_{\!\Delta} (\tau_1) |\Psi_{\rm dS} \ra
\eea
between two local operators is given by the Green's  function $G(x_2,x_1)$ of the associated quantum field, evaluated at the location of the observer at time $\tau_1$ and $\tau_2$. The time difference $\tau = \tau_2-\tau_1$ is equal to the geodesic distance between the two space-time points $x_1$ and $x_2$.

Adopting this observer viewpoint, we collect the following (incomplete) wish-list of desired properties of a candidate dual quantum description of de Sitter space:
\addtolength{\parskip}{-2mm}
\begin{itemize}
\addtolength{\parskip}{-2mm}
\item{The model has a dimensionless coupling $\lambda$ that governs the ratio between the Planck length $\ell_{\smpc \rm P}$ and the de Sitter radius $R_{\rm \smpc dS}$. Time resolution is finite and the energy spectrum is bounded.}
\item{The operator algebra contains a sub-algebra of `simple operators' that in the $\lambda \to 0$ limit reduces to the algebra of QFT operators acting along a geodesic in de Sitter space.}
\item{The quantum system has a maximum entropy state $\Psi_{\spc \rm dS}$. Correlation functions of simple local operators in this state look thermal with temperature $T= 1/2\pi$ in de Sitter units.}
\item{The relative entropy between a typical state $\Psi_{M}$ with energy $M$ and $\Psi_{\rm dS}$ equals the difference between the GH entropy of the corresponding Schwarzschild-dS and dS space-times.}
\item{The model encodes the quasi-normal behavior and gravitational interactions of de Sitter space. The strength of these gravitational interactions are governed by $\lambda$.}
\end{itemize}

\addtolength{\parskip}{2mm}
 
Guided by these aspirational principles and the formal links between double scaled SYK and 3D de Sitter gravity, 
we will assemble our proposed holographic dual of  low dimensional de Sitter. The model consists of a pair of two identical double scaled SYK models coupled via an equal energy constraint: the two SYK energies must be the same. This particular adaptation of the SYK model is motivated by the correspondence between the SYK collective field theory and the Chern-Simons formulation of de Sitter gravity \cite{Witten:1988hc,Witten:1989ip}. Specifically, it can been seen that the collective theory of the doubled model maps to a sum of two Liouville CFTs with combined central charge $c=26$ and that imposing the equal energy constraint amounts to gauging the Virasoro symmetry, \cite{HVtalks}  \cite{Zamolodchikov:2005fy}\cite{ustwotwo}. We will call this adaptation of the SYK model the doubled SYK model. 

Next we will introduce a natural set of physical observables that preserve the energy constraint. Using the known expression of the SYK correlation functions, we then compute their two point functions, defined as the expectation value with respect to the maximal entropy state $|\Psi_{\rm dS}\ra$. We will show that in the large $N$ limit they match with the two-point function of local quantum fields along a geodesic world-line in 3D de Sitter space time. In particular, we will find that the two point function of the Doubled SYK model can be expanded as an infinite sum (for $\tau>0$)
\bea
\label{mainresult}
\la \Psi_{\rm dS} | {\cal O}_{\!\Delta} (\tau) {\cal O}_{\!\Delta} (0) |\Psi_{\rm dS} \ra_{|\lambda \to 0} \is \sum_{{n}} 
a_+ \, e^{-i\omega^+_n \tau} 
+ \sum_{{n}}
a_- \, e^{-i\omega^-_{n}\tau} 
\eea
over both towers of quasi-normal modes \cite{Du:2004jt,Lopez-Ortega:2006aal,Sun:2020sgn} (with zero angular momentum $\ell=0$)
\bea
\begin{array}{c}{\omega^+_n=\, -2i(\Delta \spc +\spc  n)\ \  \ \ } \\[2.5mm]{\omega^-_n =  -2i(1-\Delta \spc +\spc  n)}\end{array} 
\qquad \ \ 
\qquad 4\Delta(1-\Delta) = m^2
\eea
of a particle of mass $m$ in 3D de Sitter space. 
This precise match motivates our proposal that the coupled system of SYK models provides a dual holographic description of the 2D gravity theory obtained via a circle reduction of 3D de Sitter gravity. In a separate paper \cite{ustwotwo} we will describe other tests of this proposed duality, leaving a more detailed verification to future work. Our results give positive evidence that the duality checks most of the entries on the above wishlist.

\subsection{Entropy and Energy Scales}

Before presenting our model, it may be useful to summarize the logic by which we will determine the de Sitter radius in terms of the parameters of the SYK model. 

Our holographic model is based on double scaled SYK with $N$ Majorana fermions and a $p$-th order interaction Hamiltonian. We take the large $N$ and $p$ limit, keeping $\lambda = 2p^2/N$ fixed. The energy spectrum of our model is conveniently parametrized in terms of a real angle $\theta$ via \cite{Berkooz:2018qkz}
\bea
\label{espec}
E(\theta) \is - \frac{2 \mathbb{J}\spc 
\cos \theta}{\lambda}
\eea
with ${\mathbb{J}}$  the SYK coupling and $\lambda$ taken small. 
The angle $\theta$ runs from $0$ to $\pi$ and the energy range is therefore bounded.  The spectral density  has a symmetric gaussian-like shape centered around the zero energy state $|E_0\ra$ with $E_0 = E(\theta_0) = 0$ at $\theta_0 = {\pi}/{2}$. Hence this state is a state with maximal entropy and infinite temperature. In this paper, we will mostly consider the high temperature regime where
\bea
\pi - 2\theta \equiv \pi v \ll 1, \quad 
\label{etheta}
\qquad\quad E_0 - E(\theta) \! & \! \simeq \! & \!  \frac{\mathbb{J} \pi v}{\lambda}.
\eea

Two guiding themes to our story are that we wish to interpret the maximal entropy state with energy $E=E_0$ as a candidate state for de Sitter space and the angle $\pi v$ as a geometric feature of the space-time dual to the state with energy  $E(\theta)$. Specifically, we propose that the angle $\pi v$ can be identified (possibly up to a constant factor of 2) with a deficit angle of a (circle reduction of) 3D Schwarzschild-de Sitter space produced by a localized mass $M$
\bea
\label{angles}
2\pi \alpha \! & \! \simeq \! & \!  8\pi G_N M \spc .
\eea
Equating the mass $M$ with the SYK energy difference $E_0 -E(\theta)$ gives the following relation 
\bea
\frac{1}{4G_N\!} \;  {2\pi \alpha} \!\is\! \frac{2{\mathbb J}}{\lambda} \; \spc \pi^2 v
\eea

Another important clue, derived from the match between the SYK and de Sitter two-point functions, is that an observer interacting with the SYK model will experience the maximal entropy state $|\Psi_{\rm dS}\ra$ as an environment with a finite temperature $T ={\mathbb{J}}/{2\pi}$. We propose that this temperature should be identified with the de Sitter temperature of the gravity dual 
\bea
T = \frac{\mathbb{J}}{2\pi} \! \is \! \frac{1}{2\pi R_{\rm dS}}\spc = \, T_{\rm dS}\spc .
\eea 
Using the first law of thermodynamics $\delta E = T \delta S$, the observer will thus infer that the entropy of the environment depends on the energy $E$ via
\bea
\label{sobs}
S_{\rm obs}(E_0)  - S_{\rm obs}(E) \is  \! \frac{2\pi}{\lambda}\, \pi v\spc .
\eea
We propose that \eqref{sobs} should be compared with the expression for the entropy difference between pure 3D de Sitter and a Schwarzschild-de Sitter black hole with deficit angle $2\pi \alpha$
\bea
\label{ssds}
S_{\rm \smpc dS} - S_{\rm\smpc SdS}(M)\! \is \! \frac{ \spc R_{\rm\smpc dS}}{4 G_N} \; 2\pi \alpha. 
\eea
In an upcoming paper \cite{ustwotwo} we will give evidence in support of the identification of $\pi v = 2\pi \alpha$.
Equating the two entropy differences \eqref{sobs} and \eqref{ssds} then leads to the following identification of the de 3D Sitter radius of the emergent gravity theory in terms of SYK parameters
\bea
\label{dsradius}
 \frac{R_{\rm\smpc dS}\strut\!}{G_{\rm\smpc N}\strut\!}  \! \is \!  \frac{{8\pi}{\strut}}{{\lambda}{\strut}}\, = \, \frac{4\pi N}{ p^2}
\eea

\def\darkblue{blue!85!black}
\def\darkred{red!60!black}
\def\darkgreen{green!50!black}
\def\lcyan{cyan!50!white}

\def\lgray{gray!50!white}
\section{Doubled SYK Model}
\vspace{-1.5mm}

Here we introduce the Doubled SYK model and describe its physical spectrum and observables.
The action of the doubled model is given by the sum $S \, = \,  S_\lL + S_\rR$ of two SYK actions
\bea
\label{sykdouble}
S_{\lL} \,=\, \int\!\! dt \Bigr(\sum_{i=1}^N {\mbox{\large $\frac{\mathrm i}{2}$}}\psi^{\lL}_i \dot{\psi}_i^\lL\! - H_{L}\Bigr)
\ \ \ \ \ \ & & \ \ \ \ \, S_{\rR} \,=\, \int\!\! dt \Bigr(\sum_{i=1}^N {\mbox{\large $\frac{\mathrm i}{2}$}}  \psi^\rR_i  \dot{\psi}_i^\rR  +  H_{R}\Bigr)\nonumber \\[-1mm] \\[-1mm]
H_{L} \spc = \spc {i}^{p/2}\! \sum_{i_1\ldots i_p}\! J_{\lL,i_1\ldots i_p} \psi^\lL_{i_1} \ldots \psi^\lL_{i_p},  \; \ \ & &  \ \ \; \nonumber
H_{ R} =  \, {i}^{p/2} \! \sum_{i_1\ldots i_p} \! J_{\rR,i_1\ldots i_p} \psi^\rR_{i_1} \ldots \psi^\rR_{i_p}\, .
\eea
The SYK couplings of the two SYK models are gaussian random variables with variance
\bea
\bigl\langle  (J_{\lL,i_1 \ldots i_p})^2\bigr\rangle   =  \bigl\langle  (J_{\rR,i_1\ldots i_p})^2\bigr\rangle\! \is\!  \frac{N \mathbb{J}^2 }{2p^2 \bigl(  {N\atop p}  \bigr)}.\
\eea 
For now we will remain noncommittal about whether we require the left and right couplings to be identical or allow them to be chosen independently from the same random ensemble.

Upon quantization, the majorana oscillators satisfy the commutation relations 
\bea
\{ \psi_i^\lL, \psi_j^\lL\}  =  \{ \psi_i^\rR, \psi_j^\rR\} = 2\delta_{ij}  \ \ & &  \ \{ \psi_i^\lL, \psi_j^\rR\}  =  0.
\eea
We will consider the model in the double scaling limit
\bea
N\to \infty, \ p \to \infty, \qquad q \equiv e^{-\lambda} = e^{-\frac{2p^2}{N}} = {\rm fixed}.
\eea
The SYK model is exactly soluble in this limit and exact expressions for the spectrum and correlation functions are known \cite{Cotler:2016fpe,Berkooz:2018qkz,Berkooz:2018jqr}.\footnote{Below we will consider the $\lambda \to 0$ limit which was studied recently in \cite{Goel:2023svz,Mukhametzhanov:2023tcg,Okuyama:2023bch}.}

Up to now the left and right models are fully decoupled. Indeed, since we have two separate SYK Hamiltonians $H_L$ and $H_R$, we can in principle introduce two independent notions of  time $t_L$ and $t_R$, with time flows generated by the respective Hamiltonians. The new element in our story \cite{HVtalks} is that we will couple the left and right model via the equal energy constraint 
\bea
\label{equalh}
H_L - H_R = 0.
\eea
Physically, we may view this constraint as a form of gravitational dressing, in which one of the SYK models plays the role of a physical clock that keeps track of the time flow of the partner system, thereby identifying $t_L = t_R$. The condition \eqref{equalh} is satisfied if the two SYK systems are in a thermal field double or mixed double state \cite{Verlinde:2020upt}. However, here we will not just view \eqref{equalh} as a property of a particular preferred state, but as a microscopic restriction that identifies the physical states and operators of our system. In other words, we will require that physical operators must preserve the equal energy condition. We will discuss the physical operators in subsection 2.3.

There are several ways of implementing the constraint \eqref{equalh}. One physical way is to add a suitable interaction Hamiltonian that gaps out all states with $E_\lL-E_\rR\neq 0$, c.f. \cite{Maldacena:2018lmt}. Alternatively, either as a microscopic or effective description, we can choose to enforce the constraint \eqref{equalh} by introducing a dynamical lagrange multiplier field $e$ and define the doubled SYK model as
\bea
S = \int\! dt\; \Bigl(\sum_{i} {\mbox{\large $\frac{\mathrm i}{2}$}} (\psi^{\lL}_i \dot{\psi}_i^\lL\! + \psi^{\rR}_i \dot{\psi}_i^\rR ) -
e \spc (H_L- H_R)\Bigr).
\eea
Here $e$ is time dependent and plays the role of an einbein. 
The new action $S$ then describes a topological quantum system with a local invariance under the group of time reparametrizations. Infinitesimal reparametrizations act on the three types of dynamical fields $\psi_i^\lL$, $\psi_i^\rR$ and $e$ via
\bea
\label{diffeosym}
\delta_\xi \psi_i^\lL = \xi \dot{\psi}{}_i^\lL, \qquad 
\delta_\xi \psi_i^\rR = \xi \dot{\psi}{}_i^\rR,\qquad \delta_\xi e = \xi \dot{e}  + \dot\xi {e}
\eea 
with $\xi$ an arbitrary infinitesimal function of time.
The functional integral thus has a gauge redundancy. Following the familiar treatment of a relativistic point particle \cite{vanHolten:1995ds}, we can eliminate this redundancy by imposing the gauge condition that the einbein $e$ is a time-independent constant. Or if we want to be more fancy, we can introduce a corresponding gauge fixing action $S_{\rm gf} =  \int\! dt \,\bigl( f \dot{e} + \dot{b} \spc \dot{c} \bigr)$ where $f$ is a bosonic lagrange multiplier imposing the gauge condition $\dot{e} = 0$, and $b$ and $c$ are anti-commuting ghost variables \cite{vanHolten:1995ds}. The total gauge fixed action then has a BRST invariance generated by the conserved and nilpotent BRST charge $
Q = c (H_L -H_R) + f \pi_b,$ with $\pi_b = \dot{c}$
the canonical momentum dual to $b$. We impose that physical states and observables satisfy the physical state conditions
\bea
\label{brstinv}
Q|{\rm \smpc phys \smpc}\ra \spc =\, 0 \ \ &&  \ \ [Q,{\cal O}^{\rm \smpc phys \smpc}]\spc =\, 0\, .
\eea
We will study the solutions to these conditions below.

\smallskip

\subsection{Physical States}

\vspace{-1mm}

For states  at zero ghost number, the physical state condition \eqref{brstinv} reduces to
\bea
\label{zerotwo}
(H_L \! - H_R)|{\rm \smpc phys \smpc} \ra\!  \is\! 0.
\eea
This equal energy constraint \eqref{zerotwo} only has solutions under the special circumstance that the two SYK hamiltonians have exactly equal random couplings $
J_{\lL,i_1 \smpc . . . \smpc i_p} = J_{\rR,i_1 \smpc . . . \smpc i_p}.$
In this special case, every energy eigenstate of the right SYK model is paired up with a unique energy eigenstate of the left SYK model. We will denote such paired energy eigenstates as
\bea
\label{epair}
|E \ra \!\is\! | \spc E\smpc \ra_{\! L} \, | \smpc E\smpc \ra_{\! R}
\eea
The paired energy eigenstates satisfy the physical state condition \eqref{zerotwo} and are eigen states of the physical Hamiltonian
\bea
\label{diffham}
H |E\ra = E |E\smpc \ra, \qquad {\rm with} \qquad H = \frac 1 2 ( H_{L}+ H_R)
\eea
We interpret the time evolution generated by $H$ as the physical time of the Doubled SYK model.

The requirement that the SYK models have exactly equal random couplings leads to correlations between the L and R systems in disorder averaged quantities. In this following, we will assume that these correlations will not affect the quantities of interest in this paper. As a justification, we can either rely on the presumption that, for the quantities of interest, the two SYK models are independently self-averaging and that the correlation functions of the paired model factorize into a product of disorder averaged correlation functions. Alternatively, we could choose to work with a doubled model in which the random couplings are in fact different $J_{\lL,i_1 \smpc . . \smpc i_p} \neq J_{\rR,i_1 \smpc . . \smpc i_p}$, but taken from the same gaussian random ensemble. In the latter case, it remains true that for every energy eigenstate of the SYK$_R$ model, we can always find an energy eigenstate of the SYK$_L$ model with almost exactly equal energy. By construction we then choose the nearest one and pair them up as in \eqref{epair}. The typical energy difference between these two energy states is inversely proportional to the level density of a single SYK model. In this set-up, we can thus impose the equal energy condition \eqref{zerotwo} up to accuracy $\Delta E$ of order~$e^{-S(E)}$. For now, we leave it open which of the two types of equal energy constraints we consider.

\begin{figure}[t]
\begin{center}
\begin{tikzpicture}[scale=1.1]
  \draw[->] (-4,0) -- (0,0) node[below] {$E=0$} -- (4,0) node[right] {$E$} ;
  \draw[->] (0,0) -- (0,3) node[right] {$\rho(E)$};
  \draw[thick, blue] (0,0) plot[domain=-4:4, samples=100] (\x,{.003*exp(-\x*\x/2)*sinh(2*sqrt(4-\x))*sinh(2*sqrt(4+\x))});
\end{tikzpicture}
\end{center}
\vspace{-2.5mm}
\label{fig:spectrum}
\caption{The spectral density of the Doubled SYK model looks like a gaussian centered on $E= 0$. The energy eigenstate $|E_0\ra$ with $E_0=0$ is the state of maximal entropy $S(E_0) = \log \rho(E_0)$.}
\end{figure}
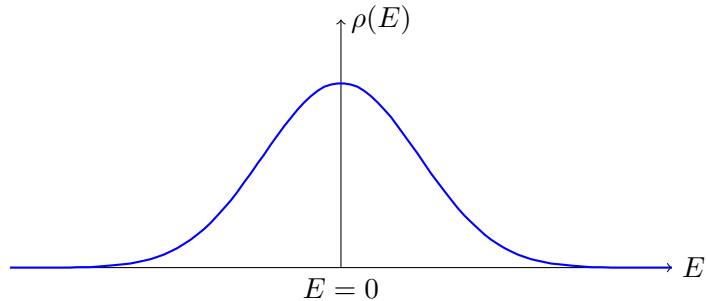

\subsection{Energy Spectrum}

\vspace{-1mm} 

As argued above, there is one solution to \eqref{zerotwo} for every energy eigen state $|E\ra_L$. The spectral density of the Doubled SYK model thus equals that of a single SYK model. The exact form of the energy spectrum of double scaled SYK is explicitly known \cite{Cotler:2016fpe,Berkooz:2018jqr,Berkooz:2018qkz}. It is most conveniently parametrized in terms of an angle variable $\theta = \lambda s$ that runs from 0 to $\pi$ via \cite{Berkooz:2018jqr,Berkooz:2018qkz} \footnote{Here and from now on we use units so that the SYK coupling set equal to $\mathbb{J} = 1$.}
\bea
\label{espectrum}
E \!\is\! \frac{-2\cos \lambda s }{\sqrt{\lambda (1-q)}}, \qquad \qquad
\rho(E) \spc = \spc \rho_0 \, { \vartheta_1(2\lambda s, q) } 
\eea
where $\vartheta_1(\nu,q)$ denotes the Jacobi theta function defined in Appendix A. 
In the limit of small $\lambda$ or $q = e^{-\lambda}$ close to 1, we can approximate (absorbing all constant factors in the pre-factor $\rho_0$)
\bea
\label{rholim}
\rho(E){}_{| \lambda \to 0} 
\!& \simeq \! & \!  \rho_0\,
\, e^{-{2\lambda}(s-s_0)^2} 
\sinh\bigl(2\pi  s\bigr) \sinh\bigl({2\pi (2s_0 -s)}\bigr) ,\qquad s_0 = \frac{\pi}{2\lambda}
\eea
As shown in figure 1, the spectral density of the double scaled SYK model takes the approximate gaussian form with a maximum at $E_0=0$ and exhibits a symmetry under $E \to - E$.  The entropy attains a maximum value equal to $S_0 = \log \rho(0)$ at $E_0=0$. 

The special energy eigenstate 
\bea
|\Psi_{\rm dS} \ra \is | E_0\ra, \qquad \qquad E_0 = 0
\eea
is the maximal entropy state, in the sense that it lies at the maximum of the spectral density \eqref{rholim}. In the following, we will identify $|\Psi_{\rm dS} \ra$ with the de Sitter vacuum state. A bit more generally, we can choose the de Sitter vacuum to be described by a macro-canonical density matrix \cite{Verlinde:2020upt} 
\bea
\rho_{\rm dS} \is e^{-S_0} \sum_{{|E-E_0|<\delta}} |E\ra \la E| 
\eea
Here the sum runs over all energy eigenstates very close to the maximal entropy state.  $S_0$ and $\delta$ are tuned such that $\tr(\rho_{\rm dS})=1$. 

For the computations we will do in this paper, it will not matter which of the above two definitions of the de Sitter vacuum state we take. Both types of states $|\Phi_{\rm dS}\ra$ and $\rho_{\rm dS}$ define infinite temperature states with energy $E_0$. It will be important, however, that the de Sitter vacuum is defined as a micro-canonical state with a definite energy $E=E_0$ and not as a unit density matrix defined on the whole spectrum of SYK states.

\subsection{Physical Observables}

\vspace{-1.5mm}

Next we turn to the physical operators of the Doubled SYK model. These must commute with the BRST charge, or equivalently, must preserve the equal energy condition \eqref{zerotwo} 
\bea
\label{physopco}
 \bigl[Q, {\cal O}^{\rm \smpc phys \smpc} \bigr] \!\is\! 
\bigl[H_L\!\spc-
H_R, {\cal O}^{\rm \smpc phys \smpc} \bigr] =  0.\ 
\eea
Hence physical operators are invariant under the time flow generated by the difference Hamiltonian. 

Local SYK operators of the $L$ and $R$ model are defined via
\bea
\label{scalingo}
{\cal O}^{\lL}_{\Delta_l} =  i^{\spc l/2}\!\! \sum_{i_1,\ldots,i_{l}}\! J^{{\cal O}_{L}}_{i_1i_2 . . i_{l}}\psi_{i_1}\psi_{i_2} . . . \psi_{i_{l}}\qquad \quad
{\cal O}^{\rR}_{\Delta_r} \!\! \is \!  i^{r/2}\! \!\sum_{i_1,\ldots,i_{r}}\! J^{{\cal O}_{R}}_{i_1i_2 . . i_{r}}\psi_{i_1}\psi_{i_2} . . . \psi_{i_{r}}
\eea
with $J^{\cal O}_{i_1,\ldots,i_{r}}$ a suitably normalized gaussian random parameters. In the double scaling limit, the number $r =\Delta p$ of SYK fermions in the product is identified with the scaling dimension at low temperature. Here we will be interested in the high temperature regime. However, we will see that labeling the local SYK operators with $\Delta$ remains useful. We will call operators of the form \eqref{scalingo} scaling operators.

To find the physical operators that satisfy \eqref{physopco}, let us first start in the conformal low temperature regime and consider operators given by the integral of the product of two scaling operators ${\cal O}^\lL_{\Delta_\lL}\!$ and ${\cal O}^\rR_{\Delta_\rR}\!$ over the time flow generated by the contraint Hamiltonian~$H_L-H_R$
\bea
\label{ophys}
{\cal O}^{\rm \smpc phys \smpc}_{\!\Delta_\lL,\Delta_\rR}\! \is\!  \int\!  d t\, e(t)^{1-\Delta_\lL-\Delta_\rR}\, {\cal O}^\lL_{\Delta_\lL}\!(t)\spc {\cal O}^\rR_{\Delta_\rR}\!\spc(-t)\ 
\eea
More generally, in the integrand on the right-hand side, it is more useful to replace the product $J^{{\cal O}_{L}}_{i_1. . . \spc i_{l}} J^{{\cal O}_{R}}_{i_1. . . \spc i_{r}}$ of independent gaussian random couplings by a single gaussian random coupling $J^{{{}_{{\cal O}{\rm phys}}}}_{{}_{i_1i_2\ldots i_{l+r}}}$. We further included a power of the einbein to absorb the scale dimension of the operators and ensure that the integral is time diffeomorphism invariant. It is straightforward to verify that in the conformal regime \eqref{ophys} define BRST invariant physical operators. From now on we will restrict to the natural subclass of Weyl invariant operators for which $\Delta_\lL + \Delta_\rR =1$. 

Next we introduce a dependence on the time flow $\tau$  generated by the physical Hamiltonian $H$
\bea
{\cal O}^{\rm phys}_{\!\Delta}(\tau)\!&\!\equiv \!&\! e^{i\tau H} \spc {\cal O}^{\rm \smpc phys \smpc}_{1-\Delta,\Delta}\spc  e^{-i \tau H} \, = \,
\int\! dt \, {\cal O}^\lL_{1-\Delta}(t) {\cal O}^\rR_{\Delta}(\tau-t)
\eea
We will make the reasonable assumption that these operators remain physical in the high temperature regime. Our results below provide evidence that this is indeed the case.

Taking the product of the two operators ${\cal O}^\lL_{1-\Delta}$ and ${\cal O}_\Delta^\rR$  and performing the integral over time amounts to a form of gravitational dressing.
It is indeed instructive to compare the above set up with the abstract construction of \cite{Chandrasekaran:2022cip} of an operator algebra for de Sitter space. An essential step in the story in \cite{Chandrasekaran:2022cip} is the introduction of an observer with a clock. As mentioned, in our Doubled SYK model we can view one of the two SYK systems as providing a clock for the other SYK system, and vice versa. The constraint $H_L-H_R=0$ ensures that both systems experience the same notion of time generated by the physical Hamiltonian $H$. 

There are some important differences between our situation and the set up considered in \cite{Chandrasekaran:2022cip}. We do not start from a type III von Neumann algebra. Rather, the operator algebra in our original SYK model, before taking the double scaling limit, is a type I algebra. After taking the double scaling limit it shares the elements of a type II algebra \cite{Bozejko:1996yv,sniady2004factoriality,Ricard2005}. Our initial set up contains some ingredients of a crossed product, with two independent Hamiltonians and the constraint that physical operators must commute with the difference of the two. Our philosophy is, in a way, the reverse of \cite{Chandrasekaran:2022cip}. We start from a microscopic theory and adopt the premise that it provides a dual description of de Sitter space with gravity. To test the proposal, we aim to show that in the semi-classical $\lambda\to 0$ limit its correlation functions match those of QFT in de Sitter space. In the latter type III algebra setting, the de Sitter vacuum is known to satisfy the KMS property with finite inverse de Sitter temperature $\beta_{\rm dS}$. We will have to explain how this property arises from the SYK model in the maximal entropy state $|\Psi_{\rm dS}\ra$ with infinite temperature. 

\def\iseqto{ \!&\!\equiv\!&\! }

With these considerations in mind, we now introduce two types of physical operators, each shifted in opposite direction by an imaginary time shift by $i\beta_{\rm dS}/4$ 
\bea
{\cal O}^{+}_{\!\Delta}(\tau) \!&\!\equiv \!&\! e^{\beta_{\rm dS} H/4} \spc {\cal O}^{\rm \smpc phys \smpc}_{\Delta}(\tau) \spc  e^{-\beta_{\rm dS}H/4} \, =\,  {\cal O}^{\rm \smpc phys \smpc}_{\Delta}\bigl(\tau- \textstyle \mbox{\Large $\frac {i\beta_{\rm dS_{\,\!}}\!} 4$}\bigr) 
\nonumber\\[-1mm]\\[-1mm]
{\cal O}^{-}_{\!\Delta}(\tau) \!&\!\equiv \!&\! e^{-\beta_{\rm dS}H/4}\spc  {\cal O}^{\rm \smpc phys \smpc}_{\Delta}(\tau) \spc  e^{{\beta_{\rm dS}H/4}}  \, =\,  {\cal O}^{\rm \smpc phys \smpc}_{\Delta}\bigl(\tau +\textstyle \mbox{\Large $\frac {i\beta_{\rm dS_{\,\!}}\!} 4$}\bigr) 
\nonumber
\eea
Here $\beta_{\rm dS}$ is a free parameter for now, that will be adjusted later on to match the KMS property of the correlation functions in the de Sitter vacuum state $|\Psi_{\rm dS}\rangle$. The $\beta_{\rm dS}/4$ shift amounts to a 90$^\circ$ rotation along the thermal circle. So geometrically, we can think of the ${\cal O}^+_{\!\Delta}$ and ${\cal O}^-_{\!\Delta}$ as the operators that sit on opposite sides of the half thermal circle associated with the thermofield double state $\rho_{\rm thermal}^{1/2} = e^{-\beta_{\rm dS} H/2}$. The two types of operators are thus related via an imaginary time shift 
\bea
\label{orels}
{\cal O}^-_{\!\Delta}(\tau) \!\is\!
{\cal O}^+_{\!\Delta}\bigl(\tau+ \mbox{\Large $\frac {i\beta_{\rm dS_{\,\!}}\!} 2$}\spc\bigr).
\eea
 Note that this relation implies that correlation functions of, say, ${\cal O}^-_{\!\Delta}$ can be re-expressed via analytic continuation in terms of correlation functions of ${\cal O}^+_{\!\Delta}$, and vice versa.

Let ${\cal P}$ denotes the parity operator that interchanges the left and right SYK model and   ${\cal T}$ the time-reversal operator, the anti-linear operator that reverses the timeflow ${\cal T} e^{iH\tau} =  e^{-iH\tau}{\cal T}$. Then
\bea
{\cal P} {\cal O}^+_{\!\Delta}(\tau){\cal P} \! \is \! {\cal O}^+_{1 - \Delta}(\tau),  \qquad
\qquad {\cal T} {\cal O}^+_{\!\Delta}(\tau){\cal T} \, = \,  {\cal O}^+_{\Delta} (-\tau).
\eea
As we will see in the next section, the holographic dictionary will identify both types of SYK operators with local field operators that emit and absorb local excitations at the antipodal north and south pole points of de Sitter space. 

\smallskip

\section{Two-Point Function}

\vspace{-.5mm}

We now turn to our main object of study, the $2\times 2$ matrix of two-point functions of the physical operators ${\cal O}^\pm_{\!\Delta}(\tau)$ 
of the Doubled SYK model evaluated between two maximal entropy states $|\Psi_{\rm dS}\ra$
\bea
\label{gabtwo}
\qquad \ G^{\smpc ab}_{\! \Delta}(\tau_2,\tau_1) \!\is\! \bigl\la \Psi_{\rm dS} \bigl| \spc {\cal O}^{\smpc a}_{\!\Delta} (\tau_2) \spc {\cal O}^{\smpc b}_{\!\Delta} (\tau_1)\spc  \bigr|\Psi_{\rm dS} \bigr\ra , \qquad a,b = \pm
\eea
We will first obtain an exact expression for this two point function at finite $\lambda$. We then take the $q\to 1$, $\lambda \to 0$ limit and show that it reduces to the Green's function of a massive complex scalar field in 3D de Sitter space. As a corollary, we find that the spectral decomposition of the Green's functions have poles at the $\ell=0$ quasi-normal mode frequencies of 3D de Sitter space. The appearance of the two towers of quasi-normal modes is directly linked to the fact that the physical operators are the product of an operator of scale dimension $\Delta$ and $1-\Delta$.

\def\Eo{\mbox{$E_1$}}
\def\E{\mbox{$E$}}
\def\Et{\mbox{$E_2$}}

\smallskip

\subsection{Two Point Function in Doubled SYK}

\vspace{-1mm}

The calculation of the two point functions is straightforward, since the double scaled SYK correlators are explicitly known \cite{Berkooz:2018qkz}. Since the two types of Green's functions are related via
\bea
\label{fourgs}
G^{-+}_{\! \Delta}(\tau_2,\tau_1) 
\! \is\!\textstyle G^{++}_{\! \Delta}\bigl(\tau_2+ \mbox{\large $\frac {i\beta_{\rm dS}\!} 2$},\tau_1 \spc\bigr),\ 
\eea
 it suffices to compute just one of them, say (here $\tau = \tau_2-\tau_1$)
\bea
\label{spectralgreeno}
G^{++}_{\! \Delta}(\tau) \! \is\! \int \! {dE} \spc G^{++}_{\! \Delta}(E)  \spc e^{-i(E-E_0)\tau}.
\eea
The function $G^{++}_{\! \Delta}(E)$ is obtained by inserting a complete set of intermediate energy states 
\bea
\label{spectralgreent}
G^{++}_{\! \Delta}(E) \is \rho(E)\spc \bigl|\la E_0  | {\cal O}^+_{\!\Delta} | E\ra \bigr|^2 
\eea
where ${\cal O}^+_{\!\Delta} = {\cal O}^+_{\!\Delta}(0)$ is the time-independent Schr\"odinger operator. 
For the rest of this subsection, we will drop the $+$ superindices.

The explicit formula for the spectral density $\rho(E)$ is given in section 2.2.  The matrix element of scaling operators in \eqref{spectralgreent} can be expressed in terms of q-deformed gamma functions as 
\bea
\label{matrixelt}
\bigl\la E_0  \bigl| {\cal O}_{\!\Delta} \bigl| E\bigr\ra \is  \bigl\la E_0  \bigl| {\cal O}^L_{1-\Delta}\bigl| E\bigr\ra_{L}\,  \bigl\la E_0  \bigl| {\cal O}^R_{\Delta}\bigl| E\bigr\ra_{R}\nonumber\\[-1mm]\label{qgammax} \\[-1mm]\nonumber
\is \sqrt{\frac{\Gamma_{q}(1\!-\!\Delta\pm is_0\pm is)}{\Gamma_q(2\Delta)}}
\sqrt{ \frac{\Gamma_{q}(\Delta\pm is_0\pm is)}{\Gamma_q(2\! -2\Delta)}} 
\eea
Here we used that the matrix element in the doubled theory  simply factorizes as a product of the left and right matrix elements. The definition of the q-deformed gamma function is given in Appendix A. In the $q\to 1$ limit with fixed $s$, it reduces to the ordinary gamma function. This expression is therefore particularly well suited for making the comparison with the correlation functions of (doubled) Schwarzian QM and JT gravity.

For our story, we are interested in the regime where $s$ lies in the neighborhood of $s_0 = \pi/2\lambda$. So $2s$ and $s+s_0$ both scale as $1/\lambda$. In this regime, the q-gamma functions do not reduce to ordinary gamma functions for small $\lambda$ and the connection with Schwarzian quantum mechanics is less clear.
To rewrite the above formula in a way that is suitable for our regime of interest we re-express the right-hand side of \eqref{qgammax} in terms theta functions by using the identity \cite{Askey} 
\bea 
\Gamma_{q \spc}(x) \Gamma_{q \spc}(1-x)\! \is \! \frac{iq^{1/8}(1-q)(q;q)^3}{q^{x/2}\vartheta_1(i\lambda x,q)}
\eea
with $\lambda = -\log q$.
This is the q-deformation of the Euler reflection formula ${\Gamma(x)\Gamma(1-x)}= {\pi}/{\sin \pi x}$ of classical gamma functions, and reduces to it in the $q\to 1$ limit. Using this identity, we can express the correlators of the Doubled SYK model by means of the following diagrammatic rules.

\smallskip

\subsection{Diagrammatic Rules}

\vspace{-1mm}
 
Correlation functions of the Doubled SYK model can be explicitly computed via the steps outlined above. The resulting expressions can be summarized via the following diagrammatic rules. Just as in DSSYK, the correlation functions can be represented as trivalent diagrams obtained by connecting propagators by means of three-point vertices.  The vertices represent the matrix elements of the local operators between energy eigenstates and the propagators represent the time evolution in the energy eigen basis. The propagators and vertices for the Doubled SYK model read 
\bea
\label{frulestwo}
\begin{tikzpicture}[scale=0.57, baseline={([yshift=-0.25cm]current bounding box.center)}]
\draw[thick] (-0.2,0) arc (160:20:1.63);
\draw[fill,\darkblue] (-0.2,0.0375) circle (0.1);
\draw[fill,\darkblue] (2.8,0.0375) circle (0.1);
\draw (3.4, 0) node {\footnotesize $\tau_1$};
\draw (-0.7,0) node {\footnotesize $\tau_2$};
\draw (1.25, 1.4) node {\footnotesize $s$};
\end{tikzpicture}\hspace{-2mm} \is e^{-iE \spc (\tau_2-\tau_1)}, \qquad \qquad E = -\frac{2\cos  \lambda s}{\sqrt{\lambda(1-q)}} . \\[7mm]
\begin{tikzpicture}[scale=0.83, baseline={([yshift=-0.1cm]current bounding box.center)}]
\draw[thick] (-.34,1) arc (40:-40:1.5);
\draw[fill,\darkblue] (0,0) circle (0.08);
 \draw[thick,\darkblue](-1.5,0) -- (0,0);
\draw (.3,-0.75) node {\footnotesize $\textcolor{black}{s_2}$};
\draw (.3,0.75) node {\footnotesize$\textcolor{black}{s_1}$};
\draw (-1,.3) node {\footnotesize\textcolor{\darkblue}{$\Delta$}};
\end{tikzpicture}\  \is \la E_2| {\cal O}_\Delta | E_1 \ra \; = \; \sqrt{\frac{\spc \vartheta_1  \bigl(2i\lambda\Delta \qbigrr\spc 
 \strut}{\smpc \vartheta_1  \bigl(\lambda(i\Delta \pm  s_1\pm s_2)\qbigrr}}_{\strut}
 \eea
The total expression obtained via these rules is integrated over the intermediate energies $E(s)$ with integration measure $\rho(E) dE$ given in equation \eqref{espectrum}. 

Applying the above rules to the two-point function gives
\bea
G_\Delta (\tau) \is \int\! d E \, G_{\Delta}(E) \,e^{-i(E-E_0) \tau}
\eea
\vspace{-5mm}
\bea
\label{gexact}
G_{\! \Delta}(E)  
\is\; \rho(E) \;\; \raisebox{3pt}{\begin{tikzpicture}[scale=0.6, baseline={([yshift=0cm]current bounding box.center)}]
\draw[thick] (0,0) circle (1.5);
\draw[thick,\darkblue] (-1.5,0) -- (1.5,0);
\draw[fill,\darkblue] (-1.5,0) circle (0.1);
\draw (0.05,1.88) node {\footnotesize $s$};
\draw[fill,\darkblue] (1.5,0) circle (0.1);
\draw (0.05,-1.88) node {\footnotesize $s_0$};
\draw (-0,.4) node {\footnotesize \textcolor{\darkblue}{$\Delta$}};
\end{tikzpicture}} 
\; = \; \rho_0\, \vartheta_1  \bigl(2\lambda s\qbigrr \;  \frac{\spc \vartheta_1  \bigl(2i \lambda\Delta \qbigrr\spc 
 \strut}{\smpc \vartheta_1  \bigl(\lambda(i\Delta \pm s_0\pm s )\qbigrr \strut} .
 \eea

\subsection{Semi-Classical Limit}

\vspace{-1mm}

In the $\lambda \to 0$ limit, the holographic dual of the SYK model reduces to semi-classical de Sitter gravity with freely propagating matter fields. We anticipate that in this limit the doubled SYK two point function reduces to the Green's function of a matter field propagating in de Sitter space. 

The $\vartheta_1$ function simplifies at small $\lambda$ to the expression given in \eqref{rholim}. Substituting $s = s_0 - \omega/2$ with $\omega\ll s_0$ into \eqref{gexact} and dropping overall factors that become constants at $\lambda\to 0$,
we find after a straightforward calculation that the spectral Green's function $G_{\! \Delta  }(E)$ reduces at small $\lambda$ to 
 \bea
 \label{gesemi}
 G_{\! \Delta  }(E)_{|\lambda= 0} \! \is\! \frac{ \sin \pi\mu \strut }{\spc \cosh\frac{\pi}{2} ({\omega} 
+ i \mu) \cosh\frac{\pi}{2} ({\omega} - i \mu) 
\strut} 
\eea
with $\mu = 1-2 \Delta$.  Appendix B summarizes some of the intermediate steps in the derivation of \eqref{gesemi}.

Fourier transforming gives the following result for the semi-classical two-point function in position space
\bea
\label{gppone}
G^{++}_{\! \Delta  }(\tau)_{|\lambda= 0}  \is 
 \frac{2\sinh( \mu \tau)}{\pi \sinh \tau\strut}, 
\eea
As we will see in the next section, this expression matches the antipodal Green's function of a massive scalar field in $dS_3$, defined as the two-point function between two antipodal points. We note in particular that \eqref{gppone} remains regular at the coincident point $\tau=0$ but has a singularity $\tau = i \pi$ corresponding to the configuration where both operators are placed at antipodal points.

Motivated by this match, we define the other Green's functions via \eqref{fourgs} with $\beta_{\rm dS}$ set equal to the inverse de Sitter temperature $\beta_{\rm dS} = 2\pi$. The off-diagonal Green's function then takes the form
\bea
\label{gpmtwo}
G^{-+}_{\! \Delta  }(\tau)_{|\lambda = 0} \is  \frac{2\sinh\bigl( \mu(i\pi + \tau)\bigr)}{\pi \sinh \tau {\strut}}
\eea
As we will see shortly, this expression looks identical to the Wightman function of a massive scalar particle in the Bunch-Davies vacuum in 3D de Sitter space, measured between two points separated by geodesic distance $\tau$ \cite{Bousso:2001mw}. This match between the two point function in the doubled SYK model and the Green's function of a scalar field in de Sitter space is the main result of our paper.

\def\phip{\phi_{\! {}_+  }}
\def\phim{\phi_{\! {}_-  }}
 
\section{Towards a Holographic Dual}

\vspace{-1.5mm}

In this section we introduce a proposed candidate for a holographic dual of the Doubled SYK model.  It takes the form of a deformed 2D JT gravity model with non-linear bulk potential (a closely similar model was introduced in \cite{Blommaert:2023opb})
\bea
\label{jtmodel}
S_{\rm gravity} \is \frac{1}{\lambda}{}_{\strut} \int^{}_{\! \cal M}\!\! {\textstyle\sqrt{\mbox{\small$-$}{g}}}\,\bigl(\Phi {R} - \cosh\Phi \bigr) 
\eea

\begin{figure}[t]
\begin{center}
\begin{tikzpicture}[scale=1.03]
\draw (4,1.4) node {$dS_2$};
\draw[very thick] (2,-2) -- (6,-2);
\draw[very thick] (2,2) -- (6,2);
\draw[dashed] (2,-2) -- (6,2);
\draw[thick,blue] (6,-2) -- (6,2);
\draw[thick,blue] (2,-2) -- (2,2);
\draw[dashed] (6,-2) -- (2,2);
\draw (6.3,0) node {\rotatebox{90}{\footnotesize \textcolor{blue}{$UV = \! -\spc 1$}}};
\draw (4,-2.3) node {\footnotesize $UV = 1$};
\draw (1.7,0) node {\rotatebox{90}{\footnotesize\textcolor{blue}{$UV = \! -\spc 1$}}};
\draw (4.85,-.5) node {\rotatebox{-45}{\footnotesize $U\!= 0$}};
\draw (3.15,-.5) node {\rotatebox{45}{\footnotesize $V\! =  0$}};
\draw (4,2.3) node {\footnotesize $UV = 1$};
\end{tikzpicture} 
\vspace{-3mm}
\end{center}
\caption{Penrose diagram of 2D de Sitter space. The time-like boundaries $UV = \! -\spc 1$ denote the location where $e^\Phi =0$.}
\end{figure}
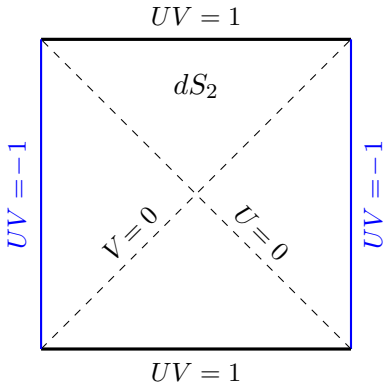

 The classical and quantum properties of deformed JT gravity models have been well studied. 
 In dilaton gravity, one has the freedom to choose different physical metrics related by Weyl rescalings. We choose our physical metric as follows
\bea
d\hat{s}{}^2 \! \is\! e^{\Phi}
\, d{s}{}^2\qquad \qquad {\rm (physical\ metric)}
\eea
In the physical Weyl frame, the deformed JT gravity theory admits a solution of the form of 2D de Sitter space with a radially varying dilaton  
\bea
\label{dsonesol}
d\hat{s}{}^2 \! \is\! -(1-\rho^2) dt^2 + \frac{d\rho^2}{1-\rho^2}, \, \qquad \quad
{\Phi} \,=\, \log \rho .
\eea
The radial coordinate $\rho$ runs over the range $0 < \rho < 1$. In Kruskal coordinates $UV = -\frac {1+\rho}{1-\rho}$, $
U/V= - e^{2t}$ the solution takes the form
\bea
\label{dstwosol}
d\hat{s}{}^2 \!\is\!  \frac{-4 dU dV}{(1-UV)^2} , \qquad\qquad e^{\Phi} \,=\,  \frac{1  +   UV}{1-UV} \spc 
\eea
The coordinate range is specified by $-1< UV< 1$. The lower boundary corresponds to the location where $e^\Phi =0$. This is where we will place our observer.

The solution \eqref{dstwosol} has a natural geometric interpretation in terms of 3D de Sitter gravity. 
We introduce the following 3D metric
\bea
\label{dsthree}
ds_{\rm 3D}^2 \!\is 
{d\hat{s}}{}_{\rm 2D}^2 + e^{2\Phi}d \varphi^2, 
\eea
with ${d\hat{s}}{}_{\rm 2D}^2$ and $e^{2\Phi}$ given in \eqref{dstwosol}  and
where $\varphi$ denotes an angular coordinate that runs from $0$ to $2\pi$. This metric describes 3D de Sitter space. This correspondence is not a coincidence and part of the reason why we consider this particular 2D deformed dilaton gravity model. In a concrete practical sense, we can view the 2D model as the circle reduction of pure 3D gravity.

\subsection{Matter Green's Function}
\vspace{-1mm}

The connection between the 2D gravity theory and 3D de Sitter gravity acquires more physical significance once we couple to matter, since matter propagates differently in 2 and 3 dimensions. In anticipating an interpretation of the 2D model as the holographic bulk dual of the Doubled SYK model, we consider massive 2D matter fields with the following 2D action 
\bea
\label{faction}
S_{\rm matter} \!\is \! -\int_{\cal M}\!\! \sqrt{\mbox{\small$-$}{\hat{g}}_2\!\!}\;\, e^{\Phi} \bigl(\hat\nabla \phi_+ \hat\nabla \phi_- +m^2\phi_+\phi_-\bigr)
\eea
This 2D matter action is the dimensional reduction of the action of a 3D massive complex scalar field. Its equations of motion take the form of the dimensional reduction of the Klein-Gordon equation in 3D de Sitter space. Plugging in the explicit form of the 2D solution for the 2D metric and dilaton in static coordinates gives the following wave equation
\bea
\label{wavee}
\Bigl(- \frac{1}{1-\rho^2} \spc \partial_t^2 +\frac{1}{\rho}\partial_\rho (1-\rho^2)\rho \partial_\rho - m^2\Bigr) \phi_\pm(\rho,t) \is 0
\eea
This is the $s$-wave reduction of the wave equation in dS$_3$. We will see momentarily why we chose the matter to be a complex rather than a real scalar field.

3D de Sitter is a hyperboloid  $
P(X,X) \spc \equiv \spc \eta_{ab} X^a X^b = 1,$  in a 4D Lorentzian embedding space with signature $\eta_{ab} = (-,+,+,+)$. Solutions to the wave equation \eqref{wavee} can be conveniently specified in the embedding space formalism as harmonic functions defined on the full embedding space with specified scale dimension $\Delta_\pm$ given by the two solutions to the equation $4\Delta(1-\Delta) = m^2$
\bea
\label{embedwave}
(X^a\partial_a - 2\Delta_\pm)\phi_\pm \!\is\! 0, \qquad \eta^{ab}\partial_a \partial_b \phi_\pm\spc =\spc  0
\eea

Our main object of interest is the Green's function $G(x_1,x_2)$ associated with the wave equation \eqref{wavee} between two timelike separated points $x_1$ and $x_2$ located on the world-line of the observer located at $\rho =0$. When lifted to 3D, the observer world-line becomes a geodesic in dS$_3$.  The Green's function $G(x_1,x_2)$ can also be lifted to 3D and extended to the embedding space, where it solves \eqref{embedwave} with a delta-function source. De Sitter space is a maximally symmetric space and $G$ thus depends only on the geodesic distance $\tau(x_1,x_2)$ between the two points. This geodesic distance is conveniently specified in terms of the embedding space coordinates via
\bea
P(x_1,x_2) \spc \equiv \spc \eta_{ab} X_1^a X_2^b\! \is \! \cosh\bigl( \tau(x_1,x_2)\bigr)
\eea 
As a function of $P$, the Green's function 
is given by \cite{Bousso:2001mw}
\bea
G(x_1,x_2) \! \is\!  \; \frac{\Gamma(h_+)\Gamma(h_-)}{4\pi^2} \; {}_{2_{\strut}}F_1\Bigl(h_+,h_- ;  \frac 3 2 , \frac{1 + P(x_1,x_2)}{2}\Bigr)^{\strut}
\eea
with  $h_\pm = 2\Delta_\pm$. A more practical form of the Green's function is found by noting that the wave equation for $G$ can be rewritten in terms of the geodesic distance $\tau = \tau(x_1,x_2)$ as follows
\bea
G(\tau) \, = \,  \frac{\chi(\tau)}{\sinh \tau},  \qquad  \qquad \bigl(\partial_\tau^2 - \mu^2 \bigr)\spc \chi(\tau)  \!\is \! 0 ,
\eea
with $\mu^2 = 1-m^2$. This equation has two independent solutions \cite{Bousso:2001mw}
\bea
\label{dsgreen}
G(\tau)\! \is\! \frac{2\sinh \mu (i\pi+  \tau) }{\pi \sinh \tau}^{\strut}_{\strut}, \spc 
\qquad \qquad
{G}_A(\tau) \, = \, \frac{2\sinh \mu \tau }{\pi \sinh \tau} .
\eea
The first solution is the standard Green's function of a massive particle in dS$_3$ with a singularity at $x_1=x_2$, given by the Wightman function associated to the Bunch-Davies vacuum. The second solution is the antipodal Green's function, which is regular at $x_1 = x_2$ but has a singularity when both points are antipodal $x_1= (x_{2})^\Omega$. Here $\Omega$ denotes the $\mathbb{Z}_2$ antipodal involution of de Sitter space.

\begin{figure}[t]
 \begin{center}
 \medskip
${}$~~~~~\begin{tikzpicture}[scale=1.02];
\path[draw=blue,thick, snake it] (6,1.5) arc (138:222:2.25cm);
\path[draw=blue,very thick,<-] (5.32,.06) arc (179:181:1cm);
\draw[very thick,\darkblue] (2,-2) -- (6,-2);
\draw[very thick,\darkblue] (2,2) -- (6,2);
\draw[thick,dashed,gray] (2,-2) -- (6,2);
\draw[thick,dashed,gray] (6,-2) -- (2,2);
\draw[thick,\darkred] (6,-2) -- (6,2);
\draw[thick,\darkred] (2,-2) -- (2,2);
\draw[thick,\darkred] (6,-1.5) -- (6,1.-1.5);
\draw[thick,\darkred] (2,-1.5) -- (2,1.-1.5);
\filldraw[\darkblue] (6,-1.55) circle (2pt);
\draw[thick, \darkblue] (6,1.5) circle (2pt);
\draw (6.5,-1.5) node {\rotatebox{0}{\textcolor{\darkblue}{ 
${\cal O}^+_{\tau_1}$}}};
\draw (6.5,1.5) node {\rotatebox{0}{\textcolor{\darkblue}{
${\cal O}^-_{\tau_2}$}}};
\draw (4,-2.5) node {\rotatebox{0}{\textcolor{\darkblue}{
$G^{-+}$}}};
\end{tikzpicture}~~~~~~~~~~~\begin{tikzpicture}[scale=1.02]
\draw[very thick,\darkblue] (2,-2) -- (6,-2);
\draw[very thick,\darkblue] (2,2) -- (6,2);
\draw[ thick,dashed,gray] (2,-2) -- (6,2);
\draw[ thick,dashed,gray] (6,-2) -- (2,2);
\draw[thick,\darkred] (6,-2) -- (6,2);
\draw[thick,\darkred] (2,-2) -- (2,2);
\draw[thick,\darkred] (6,-1.5) -- (6,1.-1.5);
\draw[thick,\darkred] (2,-1.5) -- (2,1.-1.5);
\draw[very thick,<-,\darkblue] (3.95,-1.45)--(4.05,-1.45);
\filldraw[\darkblue] (6,-1.5) circle (2pt);
\draw[thick, \darkblue] (2,-1.5) circle (2pt);
 \path[draw=blue,thick, snake it] (6,-1.5) -- (2,-1.5);
\draw (6.5,-1.5) node {\rotatebox{0}{\textcolor{\darkblue}{ 
${\cal O}^+_{\tau_1}$}}};
\draw (1.55,-1.5) node {\rotatebox{0}{\textcolor{\darkblue}{ 
${\cal O}^+_{\tau_2}$}}};
\draw (4,-2.5) node {\rotatebox{0}{\textcolor{\darkblue}{
$G^{++}$}}};
\end{tikzpicture}
\end{center}
\vspace{-4mm}
\caption{Schematic depiction of the two types of de Sitter Green's functions. Based on the explicit answer \eqref{gppone}-\eqref{gpmtwo} obtained from the SYK model, we interpret the ${\cal O}^+$ operator as that can emit particles from the north pole and absorb particles from the south pole and vice versa for ${\cal O}^-$. 
}
\end{figure}
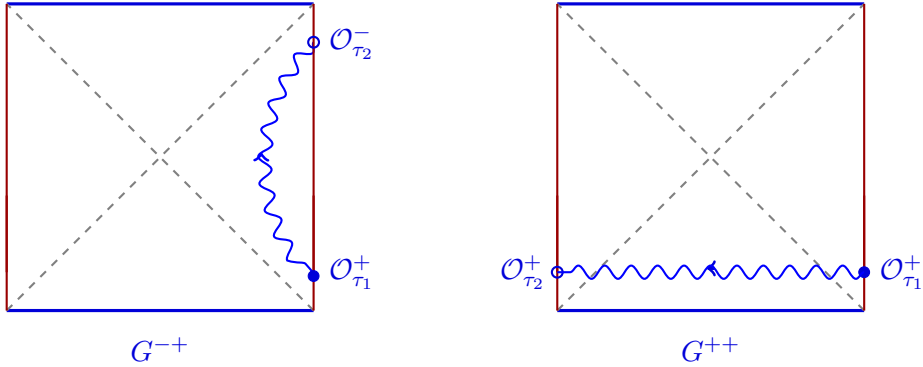

Comparing with \eqref{gppone} and \eqref{gpmtwo}, we see that there is a precise match between the 3D de Sitter Green's functions \eqref{dsgreen} and the semi-classical limit of the two-point function $G^{ab}(\tau)$ in the Doubled SYK model. 
To make this correspondence more precise, we impose the following non-local $\mathbb{Z}_2$ identification on the complex scalar field 
\bea
\label{apod}
\phi_-(x) \!\!& \! \equiv \!& \!\!  \phi_+(x^\Omega)\, .
\eea
Viewing $\phi_+$ and $\phi_-$ as each others image under charge conjugation ${\cal C}$ and the involution $\Omega$ as 
${\cal P}{\cal T}$, where ${\cal P}$ is parity and ${\cal T}$ time reversal,
this identification \eqref{apod} amounts to setting ${\cal C} {\cal P} {\cal T} = 1$. Restricted to the north pole, this identification amounts to equating 
\bea
\label{apode}
\phi_-(\tau)\! \is \!  \phi_+\bigl(\, \mbox{\large $\frac{i\beta_{\rm dS_{\,}\!}\!}{2}$}+ \tau\, \bigr),
\eea
as intended. Next we
identify the 2$\times$2 matrix of SYK two-point functions $G^{ab}_\Delta(\tau)$ in \eqref{gabtwo} with the 2$\times$2 matrix of
Green's functions of the complex scalar field defined with respect to the Bunch-Davies de Sitter vacuum.  
\bea
G^{ab}_\Delta(\tau)\! \is\! \bigl\la \Psi_{\rm dS} \bigl| \spc {\cal O}^{\smpc a}_{\!\Delta} (\tau) \spc {\cal O}^{\smpc b}_{\!\Delta} (0)\spc  \bigr|\Psi_{\rm dS} \bigr\ra =  \bigl\langle\spc \phi_a(x_1) \phi_b(x_2)\spc\bigr\rangle_{\rm dS}, \nonumber \\[-1mm]\\[-1mm]\nonumber
& &\quad \tau\spc \equiv
\tau(x_1,x_2), \quad a,b = \pm 
\eea
The two sides of the equation
match if we identify ${\cal O}_\Delta^{\spc a}$ with the corresponding charged field operator $\phi_a$ of mass $m$ acting along the worldline at the north pole of the static patch. 

To see this, we note that due to the off-diagonal form of the scalar field action \eqref{faction}, the off-diagonal correlator $G_\Delta^{+-}$ equals the standard Green's function with a pole when $x_1$ and $x_2$ are coincident. Secondly, because we imposed the condition \eqref{apod} that the two fields are each others antipodal image, we ensured that the diagonal correlator $G_\Delta^{++}$ has a pole when $x_1$ and $x_2$ are antipodal, as indicated in figure 3. This behavior of the two Green's function matches our result \eqref{gppone} and \eqref{gpmtwo} for the diagonal and off-diagonal SYK two-point functions. 
This match is a strong hint that the doubled SYK model has a dual interpretation in terms of de Sitter gravity.

The Green's functions reflect the quasi-normal mode spectrum of de Sitter space.  As noted in the introduction, they can be expanded as an infinite sum 
over both towers of quasi-normal mode frequencies 
of a particle of mass $m$ in 3D de Sitter space 
\bea
\label{mainrepeat}
G(\tau) \! \is\! \sum_{n} a_+ \, e^{-i\omega^+_n \tau} + \sum_{n} a_- \, e^{-i\omega^-_{n}\tau},  \qquad \qquad
{\omega^\pm_n=\, -2i(\Delta_\pm \spc +\spc  n)} 
\eea
with $2\Delta_\pm  = \spc 1   \pm {\mu} =  1  \pm  \sqrt{1-m^2}$. All these modes have zero angular momentum $\ell=0$ and thus descend to quasi-normal modes of the circle reduction to two dimensions. Our SYK two-point function reproduces these two towers of quasi-normal modes thanks to the fact that it is comprised of a product of two operators ${\cal O}^\rR_\Delta$ and ${\cal O}^\lL_{1-\Delta}$.

\subsection{Observing the de Sitter Temperature}
\vspace{-1mm}

Our match between the Green's functions implies that an observer interacting with the Doubled SYK model would draw the conclusion that the SYK model has a finite temperature $\beta_{\rm dS}=2\pi$ and a spectral entropy that equals the 3D Schwarzschild-de Sitter entropy. 
\bea
\label{sghtwo}
S_{\rm dS}(E_0) - S_{\rm SdS}(E) \! \is\!  2\pi (E_0-E) \qquad\qquad
\beta_{\rm dS}  = \smpc  \frac{dS_{\rm SdS}\!}{dE} \, = \, 2\pi\ 
\eea 
Here we outline how this can be made explicit, following \cite{Bousso:2001mw}.

Consider a model observer given by a single Unruh detector with a spectrum of energy eigen states  $|E_i\ra$. The detector couples to the operators ${\cal O}_\Delta^\pm$ via
\bea
S_{\rm int} \is \int\! d\tau \Bigl(X^+(\tau)\spc {\cal O}_\Delta^-(\tau) + X^-(\tau) \spc{\cal O}_\Delta^+(\tau)\Bigr)
\eea
where $X^-(\tau)$ and $X^+(\tau)$ are operators acting on the internal states of the detector that respectively raise or lower the energy of the detector. Here the integral is over the proper time along the detector worldline.  The above interaction term is time reversal symmetric.

Let $X^\pm_{ij}$ denote the matrix element of $X^\pm$ between two energy eigen states $|E_i\ra$ and $|E_j\ra$. The transition rate between two energy eigen states of the detector is then given by \cite{Bousso:2001mw}
\bea
\label{transprob}
\dot{P}(E_i\to E_j) \!\is  |X_{ij}|^2 G_\Delta(E_j-E_i) \\[3mm]
\label{gpmfour}
G_\Delta(E)\spc = \spc  \int_{-\infty}^\infty\!\!d\tau \; e^{-i E \tau}\hspace{-8mm} & &\hspace{0mm}  \la E_0\smpc |\spc {\cal O}^-_\Delta(\tau\!\spc-\! \spc i\epsilon)\spc {\cal O}^+_\Delta(0)\spc|\smpc E_0\ra
\eea
and $|X_{ij}|^2 = X^+_{ji} X^-_{ij}$. The Fourier integral $G_\Delta(E)$ of the Green's function $G_\Delta(\tau  -i\epsilon)$ along the geodesic worldline of the detector is called the detector response function. The Doubled SYK two-point function for small $\lambda$ exactly matches with Green's function of a massive scalar particle in de Sitter space. Hence the detector response function of an observer coupled to Doubled SYK model will be identical to that of an Unruh observer in de Sitter space. 

An Unruh observer in de Sitter space experiences a thermal environment. After equilibration, the detector will be in incoherent superposition of states determined by detailed balance
\bea
\label{detbal}
\frac{\dot{P}({E_i \to E_j})} {\dot{P}({E_j \to E_i})} \! \is\!  e^{-2\pi (E_j-E_i)} \, = \,  e^{S_{\rm obs} (E_i) - S_{\rm obs}(E_j)}
\eea
The ratio  on the left-hand side can be computed directly from the formula \eqref{transprob} of the transition rate in terms of the detector response function, by using the explicit form \eqref{gpmtwo} of the scalar Green's function.\footnote{We can either use the explicit expression of the spectral Green's function $G_\Delta(E)$ or apply the general argument presented in \cite{Bousso:2001mw} based on the properties of $G_\Delta(\tau)$ as an analytic function of $\tau$.}  From the detailed balance equation \eqref{detbal}, the observer would conclude that it is surrounded by an environment with a spectral density entropy ${S_{\rm obs}(E)}$ that satisfies the Gibbons-Hawking formula \eqref{sghtwo}.

\def\mfaaa{{\alpha}}

\def\mfab{\alpha}

\smallskip

\section{Concluding Remarks}

\vspace{-1mm}

We introduced and studied a new model of de Sitter holography in the form of a pair of double scaled SYK models coupled via the physical constraint that both have equal energy. De Sitter space is identified by the special state $|\Psi_{\rm dS} \ra = |E_0\ra$  with zero energy and maximal entropy. As a test of the duality, we computed the exact two-point functions $
G^{\smpc ab}_{\! \Delta}(\tau) =  \la \Psi_{\rm dS} | \spc {\cal O}^{\smpc a}_{\!\Delta} (\tau) \spc {\cal O}^{\smpc b}_{\!\Delta} (0)\spc  |\Psi_{\rm dS} \ra$
of dressed SYK operators ${\cal O}^\pm_\Delta$ that preserve the equal energy constraint.
We found that in the large $N$ limit, the two-point function matches with the Green's function of a massive complex scalar field of mass squared $m^2 = 4\Delta(1-\Delta)$ between two points $x_1$ and $x_2$ in the bulk of a 2D JT/de Sitter geometry obtained via the circle reduction of 3D de Sitter space-time.
In this correspondence, the SYK time is identified with the proper time difference $\tau_2-\tau_1$ between the two operators. The two types of Green's functions 
correspond to two different ways of emitting and absorbing a wave from the de Sitter north and south-pole, as shown in figure 3.

The match between the SYK and 3D de Sitter Green's functions is a strong hint that there exists a duality between the doubled SYK model and the circle reduction of 3D de Sitter gravity.   Bulk de Sitter Green's functions have several salient features that distinguish them from those in AdS. In particular, it exhibits two towers of regularly spaced quasi-normal modes, indicating the presence of a cosmological horizon with the characteristic de Sitter temperature $T = {\mathbb J}/2\pi = 1/2\pi R_{\rm dS}$.

\medskip

There are many open questions and directions to explore. We mention just a few of them, leaving a more detailed discussion to a future publication \cite{ustwotwo}

\medskip
\medskip

\noindent
\emph{De Sitter Entropy}

\vspace{1mm}

Our results indicate that the observable part of the entropy of the doubled SYK model, the entropy of all degrees of freedom that are accessible to an observer that can detect simple scaling operators ${\cal O}_\Delta(\tau)$, is equal to
\bea
S_{\rm obs}(E_0) - S_{\rm obs}(E)  \, \equiv \, \int_{E}^{E_0}\! \frac{dE\!}{T_{\rm obs}\!\!}   \, \is  4\pi 
(s_0-s)  
\eea
Since $s\geq 0$, we deduce the following upper bound on observable entropy in the DSSYK model
\bea
\label{sobsf}
S_{\rm obs} \leq \;  4\pi s_0 \! \is \! \frac{2\pi^2}{\lambda} \, = \, \frac{\pi^2 N}{p^2} 
\eea
As motivated in section 1.1, we anticipate that the holographic dictionary will relate DSSYK coupling with the 3D de Sitter radius via  $\lambda = 8\pi G_N/R_{\rm dS}$. This relation can be derived from the formal connections between the SYK effective field theory, Liouville conformal field theory and the Chern-Simons formulation of 3D gravity \cite{ustwotwo}. Plugging this into \eqref{sobsf} gives that
\bea
\label{sobst}
S_{\rm obs} \, \leq \, \frac{\pi R_{\rm dS}}{4 G_N} \is \frac{1}{2} S_{\rm dS}
\eea
The upper bounds \eqref{sobsf}-\eqref{sobst} are a bit surprising for two reasons. First,  \eqref{sobst} states that the observer can at most detect half of the 3D de Sitter entropy. Secondly, the upper bound \eqref{sobsf} is much smaller than the total microscopic entropy $S = N \log 2$ of the doubled SYK model. Apparently, the degrees of freedom that the observer can access and that are responsible for creating the semi-classical de Sitter cosmology are just a small fraction of the total number of microscopic SYK degrees of freedom.
This is perhaps a bit unsatisfactory: one could reasonably ask if the de Sitter gravity theory should then really be viewed as a true holographic dual of the double scaled SYK model or merely as an effective description of a small subsector. However, the same phenomenon appears in SYK/nAdS$_2$, where the JT gravity observer is also incapable of measuring the degrees of freedom responsible for the large ground state entropy $S_0$. $S_0$ appears on the 2D gravity side as an overall shift in the dilaton and the $e^{-S_0}$ expansion manifests itself as a sum over 2D space-time topologies. Our 2D JT/de Sitter gravity model also has the same freedom to shift the dilaton by a constant $S_0$\footnote{The freedom to shift the dilaton by a constant is natural from the operator algebra perspective. It has been argued that the operator algebra of the double scaled SYK model is type II \cite{Bozejko:1996yv,sniady2004factoriality,Ricard2005}. In this setting,  the von Neumann entropy of a density matrix can only be determined up to an overall additive constant.}. We expect that it can similarly be extended to include a sum over space-times with higher topology.

\medskip
\medskip

\noindent
\emph{Where is the holographic screen?}  
\nopagebreak

The correspondence between the SYK two-point function and scalar Green's function hints that the holographic screen should be thought of as located along a geodesic worldline within the emergent de Sitter space. We identify this geodesic with the north pole of the static patch. However, the doubled SYK model also includes south pole observables related to the north pole operators via an imaginary time shift by $i\beta_{\rm dS}/2$. The SYK model treats the north and south-pole points equally, and in this sense, it may be best thought of as being located symmetrically as a split screen between north and south, or as a single screen located at the midway point along the thermal half-circle connecting the north and south pode.  
The former prescription leads to an observer-centric version of holography. This is the perspective we adopted in the main text. The latter prescription places the holographic screen at a time-contour at distance $\pm i\beta_{\rm dS}/4$ away from the real time axis, as seen by observers on the north and south pole. This version of the holographic dictionary suggests that the SYK model should be viewed as located at the de Sitter horizon. This would naturally explain why it has the characteristics of an infinite temperature system \cite{Susskind:2021esx,Lin:2022nss}.  

The Doubled SYK model  incorporates a $U_{\!\mathsf q}(\mathfrak{sl}_2)$ symmetry algebra generated by three generators $K, E^+$ and $E^-$.  In the $q\to 1$ limit, the quantum algebra of the operators $K$, $E^+$ and $E^-$ reduces to a classical $SU(1,1)$ Lie algebra \cite{Berkooz:2018jqr,Lin:2023trc}. Hence we can formally exponentiate them and write operators that generate finite  $SU(1,1)$  symmetry transformations. It is tempting to identify these transformations with isometries acting on 2D de Sitter space that can move the location of the observers at the north and south pole and of the corresponding static patch. In this way, it should be possible to explore the full de Sitter space and define observables that are at arbitrary space-time points. Note that the size of the deviations from the classical symmetry generators is set by $1-q \simeq \lambda$. These corrections are quantum gravity effects that are suppressed by the ratio between the Planck scale and the de Sitter radius.

\medskip
\medskip

\noindent
\emph{Generalization to higher dimensions} 
\nopagebreak

Double scaled SYK is a 1D quantum mechanical model. According to the usual holographic paradigm, one can thus at best hope to establish a correspondence with a two-dimensional gravity theory with one extra emergent dimension. Most elements of our proposed dictionary indeed seem specific to this low dimensional setting. However, there are several reasons to be hopeful that similar ideas can give us insight into de Sitter holography in higher dimensions, including $D=4$. 

First, we have already seen indications that hint at a three-dimensional bulk gravity interpretation. Moreover, taking the point of view advocated in \cite{Witten:2023xze} and by others, a natural starting point for de Sitter holography is to first aim to construct the operator algebra along the 1D geodesic worldline of a single observer. This prescription compresses all the higher dimensional gravitational physics into the operator algebra of a 1D quantum system. Specializing to 4D holography, we should then expect that the group of space-time rotations that leave the location of the observer fixed will appear as an internal symmetry of the 1D quantum mechanics. To make progress in this direction, a promising starting point could be to look for candidate holographic duals to the near horizon dS$_2$ $\times$ S$_2$ geometry of a 4D Schwarzschild-de Sitter black hole in the Nariai limit.

\section*{Acknowledgments}

We thank Micha Berkooz, Andreas Blommaert, Scott Collier, Ping Gao, Akash Goel,  Henry Lin, Juan Maldacena, Alex Maloney, Thomas Mertens, Xiaoliang Qi, Adel Rahman, Victor Rodriguez, Eva Silverstein, Douglas Stanford, Zimo Sun, Lenny Susskind, Erik Verlinde, Jinzhao Wang, and Mengyang Zhang for useful discussions and comments. The research of VN and HV is supported by NSF grant PHY-2209997.

\appendix
\def\G{\Gamma}
\def\t{\tau}
\def\tq{\tilde q}
\def\U{\Upsilon}
\def\CP{\mathcal P}
\def\pd{\partial}
\def\Z{{\mathbb{Z}}}
\def\C{{\mathbb{C}}}
\def\CL{{\mathcal L}}
\def\CH{{\mathcal H}}
\def\Li{{\rm Li}}
\def\R{{\mathbb{R}}}

\medskip

\section{Some Useful q-Deformed Functions}

\vspace{-1mm}

In this Appendix we collect the definitions of the various special functions that appear in the main text: the q-deformed dilogarithm ${\rm Li}_2(z;q)$, the q-deformed gamma function $\Gamma_{\! q\spc}(z)$ and the Jacobi theta function $\theta_1(z,q)$.  

The quantum dilogarithm 
is defined as minus the logarithm of the $q$-Pochhammer~symbol
\bea
{\rm Li}_2(z;q) \! \is \! \sum_n \frac{z^n}{n (1-q^n)} \, = \, - \log(z;q)_{} \qquad \ \ (z;q)_{} = \prod\limits_{n=0}^{\infty} \left( 1- z q^{n} \right)
\eea
In the $q  = e^{-\lambda} \to 1$ limit it reduces to the classical Euler dilogarithm via 
\bea
{\rm Li}_2(z;q)_{|q\to 1} \! \is \! \frac{1}{\lambda}\spc {\rm Li}_2(z), \qquad {\rm Li}_2(z) = \sum_{n=1}^\infty \frac{z^2}{n^2}
\eea
The quantum dilogarithm function describes the complexified volume of a quantized ideal hyperbolic tetrahedron, see \cite{Dimofte:2011gm}. 

The $q$-analogue of the gamma function is expressed as follows \cite{Askey}
\be 
\Gamma_{q\spc}(x) \equiv \frac{(q\smpc ;q)}{(q^x;q)_{}}(1-q)^{1-x} .
\ee 
Like the usual gamma function, it satisfies a multiplicative rule under shifting the argument by~1
\bea 
\G_{q\spc} (x+1) \! \is \!  [ x ]_q \G_q(x), \quad {\rm with} \quad [n]_q \equiv \frac{1-q^n}{1-q}
\eea 
with q-deformed integer. The q-deformed gamma function $\G_q(x)$ 
reduces to the $q$-factorial when evaluated at non-negative integers and approaches the usual gamma function as $q\to 1$ from the interior of the unit disk in the complex plane $
\G_{q\spc}(x)_{|q \to 1} = \Gamma(x).$

A third class of functions used in the text are the Jacobi theta functions of the first kind 
\bea 
\vartheta_1(z,q)\!\! &\!\equiv\!&\! i \sum\limits_{n=-\infty}^{\infty} (-1)^n e^{i z\left(n-\frac{1}{2} \right)} q^{\frac 1 2 \left(n-\frac{1}{2} \right)^2}\spc= \,  \frac{ q^{\frac{1}{8}} (q;q)  } {2\sin \frac z 2  } \left( e^{\pm i z} ;q\right)_{} 
\eea 
These are periodic under shifts by 1 and quasi-periodic under shifts by $\tau$
\be \label{eq:theta1-properties}
\vartheta_1(z+2\pi,q) = - \vartheta_1(z,q) \qquad \qquad \vartheta_1(z+ \tau, q) = - e^{-i\pi\tau- i z} \vartheta_1(z,q), \qquad q \equiv e^{2\pi i \tau}
\ee
and satisfy the modular property
\bea \label{eq:theta1-s-transform}
 \vartheta_1\bigl(-{z}{/\tau}, e^{-{2\pi i}/{\tau}} \bigr) \! \is \! e^{i\pi/4} e^{i z^2/4\pi\tau} \sqrt{\tau} \, \vartheta_1\bigl(z,e^{2\pi i\t} \bigr),
\eea
The Jacobi theta-function is related to the $q$-deformed gamma function via the deformed version of the Euler reflection formula \cite{Askey} 
\bea
\Gamma_{q \spc}(x) \Gamma_{q \spc}(1-x)\! \is \! \frac{iq^{1/8}(1-q)(q;q)^3}{q^{x/2} \vartheta_1(i\lambda x,q)}
\eea

\smallskip

\section{Semi-classical Limit of the Two-Point Function}

\vspace{-1mm}
In this Appendix, we summarize some of the intermediate steps in the calculation of the semi-classical $\lambda \to 0$ limit of the exact two point function of the Doubled SYK model
\bea
\label{goneone}
G_\Delta(s) \is \; \rho_0\,  \frac{\spc \vartheta_1  \bigl(2\lambda s\qbigrr  \spc \vartheta_1  \bigl(2i \lambda\Delta \qbigrr^{\,}\spc 
 \strut}{ \smpc \vartheta_1  \bigl(\lambda(i\Delta \pm s_0\pm s )\qbigrr \large\strut}
 \eea
To simplify notation,  we will from now on drop the second argument $q$ in the theta function. The Jacobi theta functions simplify in the small $\lambda$ limit to
\bea 
 & &\ \  \vartheta_1(2\lambda y)  \, \simeq \, 
\, N_q \, e^{- 2{\lambda}(s_0-y)^2} \sinh\bigl(2\pi  y\bigr) \sinh\bigl({2\pi (2s_0 -y)}\bigr) ,\qquad s_0 \equiv \frac{\pi}{2\lambda}
\nonumber\\[-1.5mm]\label{thetalim} \\[-1.5mm]\nonumber
& & \vartheta_1\bigl(2\lambda s_0 - \lambda \omega\bigr) 
\, \simeq \, N_q \, \spc  e^{-\frac{\lambda}{2} \omega^2} \spc 
\sinh(2\pi s_0 + \pi \omega)\sinh(2\pi s_0 - \pi \omega)
\eea
with $N_q$ some $q$ dependent constant. 

To facilitate using the approximations \eqref{thetalim} in the regime of interest $s\sim s_0$, we first substitute $s=s_0 - \omega/2$ with $\omega \ll s_0$ into the expression \eqref{goneone} and write $G_\Delta(s)$ as
\bea
\label{gtwotwo}
G_\Delta(s)\! \is \! \rho_0 \, \frac{\vartheta_1\bigr(2\lambda s_0 - \lambda \omega)\bigr)}{\vartheta_1\bigl(2\lambda s_0-\lambda(\frac \omega 2 \pm i \Delta )\bigr)} \, 
\frac{\vartheta_1\bigr(2i\lambda \Delta\bigr)}{\vartheta_1\bigl(\lambda( i\Delta \pm \frac\omega 2) \bigr)}
\eea
Applying \eqref{thetalim} to the first ratio of theta functions, we find that it reduces at small $\lambda$ to a simple gaussian factor
\bea
\frac{\vartheta_1\bigr(2\lambda s_0 - \lambda \omega)\bigr)}{\vartheta_1\bigl(2\lambda s_0-\lambda(\frac\omega 2 \pm i \Delta )\bigr)}{}_{\mbox{\Large $|$}\lambda = 0}\! &\! \simeq \! & \,  \frac{e^{-\frac\lambda 4 \omega^2}}{N_q} \frac{ \sinh\bigl(2\pi s_0 \pm \pi \omega\bigr)}{\sinh\bigl(2\pi s_0 \pm \frac\pi 2 \omega \pm i \pi \Delta\bigr)} \; \simeq\;   \frac{e^{-\frac\lambda 4 \omega^2}}{N_{q}{}_{\strut}}
\eea
Similarly, we find that the second ratio of theta functions simplifies  to
\bea
 \frac{\vartheta_1\bigr(2i \lambda \Delta\bigr)^{\strut}}{\vartheta_1\bigl(\lambda(\frac\omega 2\pm i\Delta) \bigr)}{}_{\mbox{\Large $|$}\lambda = 0} \!&\! \simeq \! & \, \frac{e^{\frac \lambda 4 \omega^2 -{\pi \omega+2\pi i \Delta}} \sin(2\pi \Delta)\sinh(2\pi(2s_0 - i\Delta))}{C_q \sinh\bigl( \pi( \frac \omega 2 \pm i\Delta)\bigr)\sinh\bigl( 2\pi(2s_0- \frac \omega 4 \pm i\frac \Delta 2)\bigr) }\nonumber\\[-1mm]\\[-1mm]\nonumber \!&\! \simeq \! & \,  \frac{e^{\frac \lambda 4 \omega^2} \sin(2\pi \Delta)}{D_q\sinh\bigl( \pi( \frac \omega 2 \pm i\Delta)\bigr)}
\eea
with $C_q$ and $D_q$ some $q$ dependent constants. Plugging the above two equations into \eqref{gtwotwo}, and dropping overall $q$-dependent constants, we find that
\bea
\label{gthreethree}
G_\Delta(s)_{|\lambda = 0} \! \is \! - \frac{\sin(2\pi \Delta)}{ \sinh\bigl( \pi( \frac \omega 2 + i\Delta)\bigr)\sinh\bigl( \pi( \frac \omega 2 - i\Delta)\bigr)}
\eea
This expression for the semi-classical two-point function  equals the result \eqref{gesemi} given in section~3.3. Note that the result \eqref{gthreethree} is symmetric under reflection $\omega \to -\omega$.

\section{High $T$ Limit of Double-Scaled SYK} 

A key element in our story is that the two-point function evaluated in the maximal entropy state of the Doubled SYK model has an effective finite temperature $T ={1}/{2\pi}$. This result appears at odds with the fact that the inverse temperature computed via the SYK spectral density vanishes at the maximal  entropy point. Here we point out that a similar phenomenon takes place in a single copy of the DSSYK model.

Consider the 2-point function at inverse temperature $\beta$ between two SYK operators  of scale dimension $\Delta$. In the $q\to 1$ limit, it takes the form \cite{Maldacena:2016hyu}
\bea
\label{doubleg}
{\rm G}_\Delta(\tau_E)\is \left[ \frac{\cos \frac{\pi v}{2} }{\cos \left( \frac{\pi v}{2} \left( 1-\frac{2\tau_E }{\beta } \right) \right) } \right] ^{2\Delta}, \qquad \beta = 
\frac{\pi v}{ \cos\bigl(\frac {\pi v}{2}\bigr)}
\eea
where we have set $\mathbb{J}=1$ and $\tau_E$ is the Euclidean time.
In the low temperature limit  $v\to 1$, ${\rm G}_\Delta$ exhibits the usual scaling behavior of a two-point function of operators of dimension $\Delta$.
In the high temperature limit $v\to 0$, equation \eqref{doubleg} reduces~to a periodic function with periodicity $\beta_{\rm dS} = 2\pi$ in the imaginary time direction. Indeed, continuing to real time $\tau_E=i\tau$, we have
\bea \label{eq:large_T_2pf}
{\rm G}_\Delta(\tau)_{|v\to 0}  \! \is \left(\frac{1}{\cosh \tau}\right)^{2\Delta} , \qquad
{\rm G}_\Delta(\tau) = {\rm G}_\Delta(\tau + i \beta_{\rm dS}), \qquad \beta_{\rm dS} = {2\pi}
\eea
Note, however, that  \eqref{eq:large_T_2pf} is regular at $\tau=0$ and has a singularity at $\tau = i \beta_{\rm dS}/4 = i\pi/2$.

\begin{figure}[t]
\centering
\includegraphics[width=0.28\textwidth]{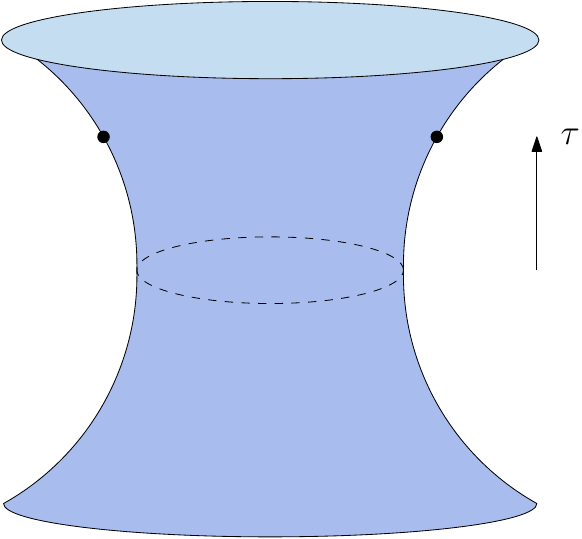}
\caption{2-point function in 2D de Sitter space for spatially separated points.}
\label{fig:ds_2pf}
\end{figure}

We can interpret \eqref{eq:large_T_2pf} as the 2-point function of CFT operators spatially separated in de Sitter space, for example for two antipodal points on the spatial sphere, one sitting in the north pole and one in the south pole as shown in Fig.\ \ref{fig:ds_2pf}. Indeed, consider any conformal field theory in two dimensional de Sitter space. The metric in global coordinates is
\bea
ds^2\! \is \! -d\tau ^2+(\cosh \tau )^2d\sigma ^2
\eea
where $\tau$ is global time, $\sigma $ is an angle going from 0 to $2\pi$, and the dS radius has been set to 1. Then, the two point function of a primary operator of dimension $\Delta$ can be calculated to be\footnote{We can map this space by a conformal transformation to a flat cylinder. We notmalize the operator such that it has a normalized 2-point function when the CFT is on a plane.}
\begin{equation}
\bigl\la {\cal O}_\Delta (\tau ,\pi ) {\cal O}_{\Delta} (\tau ,0)\bigr\rangle _{\rm dS_2} = \left(\frac{1}{2\cosh \tau}\right)^{2\Delta} .
\end{equation}
This is the same as the high temperature limit of the SYK two-point function \eqref{eq:large_T_2pf}.
Note that this correlation function is expressed in terms of the proper time of the particles.

\bibliographystyle{ssg}
\bibliography{Biblio}

\begingroup\raggedright\begin{thebibliography}{10}

\bibitem{kitaevTalks}
A.~Kitaev, ``{Talks given at the Fundamental Physics Prize Symposium and KITP
  seminars},''.
\newblock \url{https://www.youtube.com/watch?v=OQ9qN8j7EZI},
  \url{http://online.kitp.ucsb.edu/online/joint98/kitaev/},
  \url{http://online.kitp.ucsb.edu/online/entangled15/kitaev}.

\bibitem{Maldacena:2016hyu}
J.~Maldacena and D.~Stanford, ``{Remarks on the Sachdev-Ye-Kitaev model},''
  {\em Phys. Rev. D} {\bf 94} (2016), no.~10 106002,
  \href{http://xxx.lanl.gov/abs/1604.07818}{{\tt 1604.07818}}.

\bibitem{Polchinski:2016xgd}
J.~Polchinski and V.~Rosenhaus, ``{The Spectrum in the Sachdev-Ye-Kitaev
  Model},'' {\em JHEP} {\bf 04} (2016) 001,
  \href{http://xxx.lanl.gov/abs/1601.06768}{{\tt 1601.06768}}.

\bibitem{Cotler:2016fpe}
J.~S. Cotler, G.~Gur-Ari, M.~Hanada, J.~Polchinski, P.~Saad, S.~H. Shenker,
  D.~Stanford, A.~Streicher, and M.~Tezuka, ``{Black Holes and Random
  Matrices},'' {\em JHEP} {\bf 05} (2017) 118,
  \href{http://xxx.lanl.gov/abs/1611.04650}{{\tt 1611.04650}}. [Erratum: JHEP
  09, 002 (2018)].

\bibitem{Berkooz:2018jqr}
M.~Berkooz, M.~Isachenkov, V.~Narovlansky, and G.~Torrents, ``{Towards a full
  solution of the large N double-scaled SYK model},'' {\em JHEP} {\bf 03}
  (2019) 079, \href{http://xxx.lanl.gov/abs/1811.02584}{{\tt 1811.02584}}.

\bibitem{Berkooz:2018qkz}
M.~Berkooz, P.~Narayan, and J.~Simon, ``{Chord diagrams, exact correlators in
  spin glasses and black hole bulk reconstruction},'' {\em JHEP} {\bf 08}
  (2018) 192, \href{http://xxx.lanl.gov/abs/1806.04380}{{\tt 1806.04380}}.

\bibitem{Lin:2022rbf}
H.~W. Lin, ``{The bulk Hilbert space of double scaled SYK},'' {\em JHEP} {\bf
  11} (2022) 060, \href{http://xxx.lanl.gov/abs/2208.07032}{{\tt 2208.07032}}.

\bibitem{Lin:2023trc}
H.~W. Lin and D.~Stanford, ``{A symmetry algebra in double-scaled SYK},''
  \href{http://xxx.lanl.gov/abs/2307.15725}{{\tt 2307.15725}}.

\bibitem{HVtalks}
H.~Verlinde, ``{Talks given at the QGQC5 conference, UC Davis, August 2019, the
  Franqui Symposium, Brussels, November 2019, at `Quantum Gravity on Southern
  Cone', Argentina, December 2019, and `SYK models and Gauge Theory' workshop
  at Weizmann Institute, December 2019},''.

\bibitem{Susskind:2021esx}
L.~Susskind, ``{Entanglement and Chaos in De Sitter Space Holography: An SYK
  Example},'' {\em JHAP} {\bf 1} (2021), no.~1 1--22,
  \href{http://xxx.lanl.gov/abs/2109.14104}{{\tt 2109.14104}}.

\bibitem{Susskind:2022bia}
L.~Susskind, ``{De Sitter Space, Double-Scaled SYK, and the Separation of
  Scales in the Semiclassical Limit},''
  \href{http://xxx.lanl.gov/abs/2209.09999}{{\tt 2209.09999}}.

\bibitem{Susskind:2022dfz}
L.~Susskind, ``{Scrambling in Double-Scaled SYK and De Sitter Space},''
  \href{http://xxx.lanl.gov/abs/2205.00315}{{\tt 2205.00315}}.

\bibitem{Lin:2022nss}
H.~Lin and L.~Susskind, ``{Infinite Temperature's Not So Hot},''
  \href{http://xxx.lanl.gov/abs/2206.01083}{{\tt 2206.01083}}.

\bibitem{Rahman:2022jsf}
A.~A. Rahman, ``{dS JT Gravity and Double-Scaled SYK},''
  \href{http://xxx.lanl.gov/abs/2209.09997}{{\tt 2209.09997}}.

\bibitem{Klemm:2002ir}
D.~Klemm and L.~Vanzo, ``{De Sitter gravity and Liouville theory},'' {\em JHEP}
  {\bf 04} (2002) 030, \href{http://xxx.lanl.gov/abs/hep-th/0203268}{{\tt
  hep-th/0203268}}.

\bibitem{Strominger:2001pn}
A.~Strominger, ``{The dS / CFT correspondence},'' {\em JHEP} {\bf 10} (2001)
  034, \href{http://xxx.lanl.gov/abs/hep-th/0106113}{{\tt hep-th/0106113}}.

\bibitem{Alishahiha:2004md}
M.~Alishahiha, A.~Karch, E.~Silverstein, and D.~Tong, ``{The dS/dS
  correspondence},'' {\em AIP Conf. Proc.} {\bf 743} (2004), no.~1 393--409,
  \href{http://xxx.lanl.gov/abs/hep-th/0407125}{{\tt hep-th/0407125}}.

\bibitem{Banks:2006rx}
T.~Banks, B.~Fiol, and A.~Morisse, ``{Towards a quantum theory of de Sitter
  space},'' {\em JHEP} {\bf 12} (2006) 004,
  \href{http://xxx.lanl.gov/abs/hep-th/0609062}{{\tt hep-th/0609062}}.

\bibitem{Anninos:2011ui}
D.~Anninos, T.~Hartman, and A.~Strominger, ``{Higher Spin Realization of the
  dS/CFT Correspondence},'' {\em Class. Quant. Grav.} {\bf 34} (2017), no.~1
  015009, \href{http://xxx.lanl.gov/abs/1108.5735}{{\tt 1108.5735}}.

\bibitem{Anninos:2011af}
D.~Anninos, S.~A. Hartnoll, and D.~M. Hofman, ``{Static Patch Solipsism:
  Conformal Symmetry of the de Sitter Worldline},'' {\em Class. Quant. Grav.}
  {\bf 29} (2012) 075002, \href{http://xxx.lanl.gov/abs/1109.4942}{{\tt
  1109.4942}}.

\bibitem{Anninos:2017hhn}
D.~Anninos and D.~M. Hofman, ``{Infrared Realization of dS$_2$ in AdS$_2$},''
  {\em Class. Quant. Grav.} {\bf 35} (2018), no.~8 085003,
  \href{http://xxx.lanl.gov/abs/1703.04622}{{\tt 1703.04622}}.

\bibitem{Anninos:2017eib}
D.~Anninos, F.~Denef, R.~Monten, and Z.~Sun, ``{Higher Spin de Sitter Hilbert
  Space},'' {\em JHEP} {\bf 10} (2019) 071,
  \href{http://xxx.lanl.gov/abs/1711.10037}{{\tt 1711.10037}}.

\bibitem{Dong:2018cuv}
X.~Dong, E.~Silverstein, and G.~Torroba, ``{De Sitter Holography and
  Entanglement Entropy},'' {\em JHEP} {\bf 07} (2018) 050,
  \href{http://xxx.lanl.gov/abs/1804.08623}{{\tt 1804.08623}}.

\bibitem{Coleman:2021nor}
E.~Coleman, E.~A. Mazenc, V.~Shyam, E.~Silverstein, R.~M. Soni, G.~Torroba, and
  S.~Yang, ``{De Sitter microstates from $T \overline{T} $ +
  \ensuremath{\Lambda}$_{2}$ and the Hawking-Page transition},'' {\em JHEP}
  {\bf 07} (2022) 140, \href{http://xxx.lanl.gov/abs/2110.14670}{{\tt
  2110.14670}}.

\bibitem{Chandrasekaran:2022cip}
V.~Chandrasekaran, R.~Longo, G.~Penington, and E.~Witten, ``{An algebra of
  observables for de Sitter space},'' {\em JHEP} {\bf 02} (2023) 082,
  \href{http://xxx.lanl.gov/abs/2206.10780}{{\tt 2206.10780}}.

\bibitem{Witten:2023xze}
E.~Witten, ``{A Background Independent Algebra in Quantum Gravity},''
  \href{http://xxx.lanl.gov/abs/2308.03663}{{\tt 2308.03663}}.

\bibitem{Gibbons:1976ue}
G.~W. Gibbons and S.~W. Hawking, ``{Action Integrals and Partition Functions in
  Quantum Gravity},'' {\em Phys. Rev. D} {\bf 15} (1977) 2752--2756.

\bibitem{Jackiw:1984je}
R.~Jackiw, ``{Lower Dimensional Gravity},'' {\em Nucl. Phys.} {\bf B252} (1985)
  343--356.

\bibitem{Witten:1988hc}
E.~Witten, ``{(2+1)-Dimensional Gravity as an Exactly Soluble System},'' {\em
  Nucl. Phys. B} {\bf 311} (1988) 46.

\bibitem{Witten:1989ip}
E.~Witten, ``{Quantization of {Chern-Simons} Gauge Theory With Complex Gauge
  Group},'' {\em Commun. Math. Phys.} {\bf 137} (1991) 29--66.

\bibitem{Zamolodchikov:2005fy}
A.~B. Zamolodchikov, ``{Three-point function in the minimal Liouville
  gravity},'' {\em Theor. Math. Phys.} {\bf 142} (2005) 183--196,
  \href{http://xxx.lanl.gov/abs/hep-th/0505063}{{\tt hep-th/0505063}}.

\bibitem{ustwotwo}
V.~Narovlansky, H.~Verlinde, and M.~Zhang, ``{Double Scaled SYK and Quantum de
  Sitter Space},'' (2023, to appear).

\bibitem{Du:2004jt}
D.-P. Du, B.~Wang, and R.-K. Su, ``{Quasinormal modes in pure de Sitter
  space-times},'' {\em Phys. Rev. D} {\bf 70} (2004) 064024,
  \href{http://xxx.lanl.gov/abs/hep-th/0404047}{{\tt hep-th/0404047}}.

\bibitem{Lopez-Ortega:2006aal}
A.~Lopez-Ortega, ``{Quasinormal modes of D-dimensional de Sitter spacetime},''
  {\em Gen. Rel. Grav.} {\bf 38} (2006) 1565--1591,
  \href{http://xxx.lanl.gov/abs/gr-qc/0605027}{{\tt gr-qc/0605027}}.

\bibitem{Sun:2020sgn}
Z.~Sun, ``{Higher spin de Sitter quasinormal modes},'' {\em JHEP} {\bf 11}
  (2021) 025, \href{http://xxx.lanl.gov/abs/2010.09684}{{\tt 2010.09684}}.

\bibitem{Goel:2023svz}
A.~Goel, V.~Narovlansky, and H.~Verlinde, ``{Semiclassical geometry in
  double-scaled SYK},'' \href{http://xxx.lanl.gov/abs/2301.05732}{{\tt
  2301.05732}}.

\bibitem{Mukhametzhanov:2023tcg}
B.~Mukhametzhanov, ``{Large p SYK from chord diagrams},'' {\em JHEP} {\bf 09}
  (2023) 154, \href{http://xxx.lanl.gov/abs/2303.03474}{{\tt 2303.03474}}.

\bibitem{Okuyama:2023bch}
K.~Okuyama and K.~Suzuki, ``{Correlators of double scaled SYK at one-loop},''
  {\em JHEP} {\bf 05} (2023) 117,
  \href{http://xxx.lanl.gov/abs/2303.07552}{{\tt 2303.07552}}.

\bibitem{Verlinde:2020upt}
H.~Verlinde, ``{ER = EPR revisited: On the Entropy of an Einstein-Rosen
  Bridge},'' \href{http://xxx.lanl.gov/abs/2003.13117}{{\tt 2003.13117}}.

\bibitem{Maldacena:2018lmt}
J.~Maldacena and X.-L. Qi, ``{Eternal traversable wormhole},''
  \href{http://xxx.lanl.gov/abs/1804.00491}{{\tt 1804.00491}}.

\bibitem{vanHolten:1995ds}
J.~W. van Holten, ``{Propagators and path integrals},'' {\em Nucl. Phys. B}
  {\bf 457} (1995) 375--407, \href{http://xxx.lanl.gov/abs/hep-th/9508136}{{\tt
  hep-th/9508136}}.

\bibitem{Bozejko:1996yv}
M.~Bozejko, B.~Kummerer, and R.~Speicher, ``{Q Gaussian processes:
  Noncommutative and classical aspects},'' {\em Commun. Math. Phys.} {\bf 185}
  (1997) 129--154, \href{http://xxx.lanl.gov/abs/funct-an/9604010}{{\tt
  funct-an/9604010}}.

\bibitem{sniady2004factoriality}
P.~{\'S}niady, ``Factoriality of {B}ozejko--{S}peicher von {N}eumann
  Algebras,'' {\em Communications in mathematical physics} {\bf 246} (2004),
  no.~3 561--567.

\bibitem{Ricard2005}
E.~Ricard, ``Factoriality of q-{G}aussian von {N}eumann Algebras,'' {\em
  Communications in Mathematical Physics} {\bf 257} (2005), no.~3 659--665.

\bibitem{Askey}
R.~Askey, ``The q-Gamma and q-Beta Functions†,'' {\em Applicable Analysis}
  {\bf 8} (1978), no.~2 125--141.

\bibitem{Bousso:2001mw}
R.~Bousso, A.~Maloney, and A.~Strominger, ``{Conformal vacua and entropy in de
  Sitter space},'' {\em Phys. Rev. D} {\bf 65} (2002) 104039,
  \href{http://xxx.lanl.gov/abs/hep-th/0112218}{{\tt hep-th/0112218}}.

\bibitem{Blommaert:2023opb}
A.~Blommaert, T.~G. Mertens, and S.~Yao, ``{Dynamical actions and
  q-representation theory for double-scaled SYK},''
  \href{http://xxx.lanl.gov/abs/2306.00941}{{\tt 2306.00941}}.

\bibitem{Dimofte:2011gm}
T.~Dimofte, ``{Quantum Riemann Surfaces in Chern-Simons Theory},'' {\em Adv.
  Theor. Math. Phys.} {\bf 17} (2013), no.~3 479--599,
  \href{http://xxx.lanl.gov/abs/1102.4847}{{\tt 1102.4847}}.

\end{thebibliography}\endgroup

\end{document}